\documentstyle{mn}
\input psfig
\def\be{\begin{equation}}
\def\ee{\end{equation}}
\def\beq{\begin{eqnarray}}
\def\eeq{\end{eqnarray}}

\def\Lfour{{L_4}}
\def\Lfive{{L_5}}

\def\aplanet{a_{\rm p}}

\def\asin{{\rm asin}}
\def\spose#1{\hbox to 0pt{#1\hss}}
\def\lta{\mathrel{\spose{\lower 3pt\hbox{$\sim$}}
    \raise 2.0pt\hbox{$<$}}}
\def\gta{\mathrel{\spose{\lower 3pt\hbox{$\sim$}}
    \raise 2.0pt\hbox{$>$}}}

\def\betag{\beta_{\rm g}}
\def\ellg{\ell_{\rm g}}
\def\betah{\beta_{\rm h}}
\def\ellh{\ell_{\rm h}}
\def\magad{\Delta m_V}

\def\aV{a_{\rm V}}
\def\am{a_{\rm M}}
\def\arcsechr{{\rm arcsec\,hr}${}^{-1}$}
\def\ah{${}^{\prime\prime}$\,hr${}^{-1}$}

\def\rdash{r${}^\prime$}
\def\udash{u${}^\prime$}
\def\gdash{g${}^\prime$}
\def\zdash{z${}^\prime$}
\def\idash{i${}^\prime$}

\def\pv{p_v}
\def\km{{\rm km}}

\newif\ifAMStwofonts
\ifoldfss
  \ifCUPmtlplainloaded \else
    \NewTextAlphabet{textbfit} {cmbxti10} {}
    \NewTextAlphabet{textbfss} {cmssbx10} {}
    \NewMathAlphabet{mathbfit} {cmbxti10} {} % for math mode
    \NewMathAlphabet{mathbfss} {cmssbx10} {} %  "   "    "
  \fi
  \ifAMStwofonts
    \ifCUPmtlplainloaded \else
      \NewSymbolFont{upmath} {eurm10}
      \NewSymbolFont{AMSa} {msam10}
      \NewMathSymbol{\upi}     {0}{upmath}{19}
      \NewMathSymbol{\umu}     {0}{upmath}{16}
      \NewMathSymbol{\upartial}{0}{upmath}{40}
      \NewMathSymbol{\leqslant}{3}{AMSa}{36}
      \NewMathSymbol{\geqslant}{3}{AMSa}{3E}

       \let\le=\leqslant
       
    \fi
  \fi
\fi % End of OFSS

\ifnfssone
  \newmathalphabet{\mathit}
  \addtoversion{normal}{\mathit}{cmr}{m}{it}
  \addtoversion{bold}{\mathit}{cmr}{bx}{it}
  \newmathalphabet{\mathbfit} % math mode version of \textbfit{..}
  \addtoversion{normal}{\mathbfit}{cmr}{bx}{it}
  \addtoversion{bold}{\mathbfit}{cmr}{bx}{it}
  \newmathalphabet{\mathbfss} % math mode version of \textbfss{..}
  \addtoversion{normal}{\mathbfss}{cmss}{bx}{n}
  \addtoversion{bold}{\mathbfss}{cmss}{bx}{n}
  \ifAMStwofonts
    \ifCUPmtlplainloaded \else
      \UseAMStwoboldmath
      \makeatletter
      \new@mathgroup\upmath@group
      \define@mathgroup\mv@normal\upmath@group{eur}{m}{n}
      \define@mathgroup\mv@bold\upmath@group{eur}{b}{n}
      \edef\UPM{\hexnumber\upmath@group}
      \new@mathgroup\amsa@group
      \define@mathgroup\mv@normal\amsa@group{msa}{m}{n}
      \define@mathgroup\mv@bold\amsa@group{msa}{m}{n}
      \edef\AMSa{\hexnumber\amsa@group}
      \makeatother
      \mathchardef\upi="0\UPM19
      \mathchardef\umu="0\UPM16
      \mathchardef\upartial="0\UPM40
      \mathchardef\leqslant="3\AMSa36
      \mathchardef\geqslant="3\AMSa3E

       \let\le=\leqslant

    \fi
  \fi
\fi % End of NFSS release 1

\ifnfsstwo
  \DeclareMathAlphabet{\mathbfit}{OT1}{cmr}{bx}{it}
  \SetMathAlphabet\mathbfit{bold}{OT1}{cmr}{bx}{it}
  \DeclareMathAlphabet{\mathbfss}{OT1}{cmss}{bx}{n}
  \SetMathAlphabet\mathbfss{bold}{OT1}{cmss}{bx}{n}
  \ifAMStwofonts
    \ifCUPmtlplainloaded \else
      \DeclareSymbolFont{UPM}{U}{eur}{m}{n}
      \SetSymbolFont{UPM}{bold}{U}{eur}{b}{n}
      \DeclareSymbolFont{AMSa}{U}{msa}{m}{n}
      \DeclareMathSymbol{\upi}{0}{UPM}{"19}
      \DeclareMathSymbol{\umu}{0}{UPM}{"16}
      \DeclareMathSymbol{\upartial}{0}{UPM}{"40}
      \DeclareMathSymbol{\leqslant}{3}{AMSa}{"36}
      \DeclareMathSymbol{\geqslant}{3}{AMSa}{"3E}

       \let\le=\leqslant

    \fi
  \fi
\fi % End of NFSS release 2

\ifCUPmtlplainloaded \else
  \ifAMStwofonts \else % If no AMS fonts
    \def\upi{\pi}
    \def\umu{\mu}
    \def\upartial{\partial}
  \fi
\fi

\input psfig
\def\spose#1{\hbox to 0pt{#1\hss}}
\def\lta{\mathrel{\spose{\lower 3pt\hbox{$\sim$}}
    \raise 2.0pt\hbox{$<$}}}
\def\gta{\mathrel{\spose{\lower 3pt\hbox{$\sim$}}
    \raise 2.0pt\hbox{$>$}}}

\title{Asteroids in the Inner Solar System II -- Observable Properties}
\author[N.W. Evans and S.A. Tabachnik]
       {N.W. Evans$^1$ and S.A. Tabachnik$^{1,2}$\\
        $^1$ Theoretical Physics, 1 Keble Rd, Oxford, OX1 3NP\\
        $^2$ Princeton University Observatory, Princeton, NJ
        08544-1001, USA}
\date{}

\begin{document}

\maketitle

\label{firstpage}

\begin{abstract}
This paper presents synthetic observations of long-lived, coorbiting
asteroids of Mercury, Venus, the Earth and Mars. Our sample is
constructed by taking the limiting semimajor axes, differential
longitudes and inclinations for long-lived stability provided by
simulations. The intervals are randomly populated with values to
create initial conditions. These orbits are re-simulated to check that
they are stable and then re-sampled every 2.5 years for 1 million
years. The Mercurian sample contains only horseshoe orbits, the
Martian sample only tadpoles.

For both Venus and the Earth, the greatest concentration of objects on
the sky occurs close to the classical Lagrange points at heliocentric
ecliptic longitudes of $60^\circ$ and $300^\circ$. The distributions
are broad especially if horseshoes are present in the sample. The
full-width half maximum (FWHM) in heliocentric longitude for Venus is
$325^\circ$ and for the Earth is $328^\circ$. The mean and most common
velocity of these coorbiting satellites coincides with the mean motion
of the parent planet, but again the spread is wide with a FWHM for
Venus of $27.8$\ah\ and for the Earth of $21.0$\ah.  For Mars, the
greatest concentration on the sky occurs at heliocentric ecliptic
latitudes of $\pm 12^\circ$.  The peak of the velocity distribution
occurs at $65$\ah, significantly less than the Martian mean motion,
while its FWHM is $32.3$\ah. The case of Mercury is the hardest of
all, as the greatest concentrations occur at heliocentric longitudes
of $16.0^\circ$ and $348.5^\circ$ and so are different from the
classical values. The fluctuating eccentricity of Mercury means that
these objects can have velocities exceeding $1000$\ah, although the
most common velocity is $459$\ah, which is much less than the
Mercurian mean motion.

A variety of search strategies are discussed, including wide-field CCD
imaging, space satellites such as {\it The Global Astrometry
Interferometer for Astrophysics} (GAIA), ground-based surveys like
{\it The Sloan Digital Sky Survey} (SDSS), as well as infrared cameras
and space-borne coronagraphs.

\end{abstract}


%
% Comment this in or out for referee mode 
%\baselineskip 15pt
%

%
\def\be{\begin{equation}}
\def\ee{\end{equation}}

\section{Introduction}

\noindent
Even though the inner Solar System is seemingly very empty of
asteroids, the suspicion is that this may be caused by the
observational awkwardness of locating fast moving objects at small
solar elongations.  Observations of the Lagrange points of Mercury,
Venus and the Earth can only be obtained in the mornings or evenings,
and are hindered by high airmasses (e.g., Whiteley \& Tholen 1998).

There have been sporadic searches for Trojans of the Earth and Mars
before. With the exception of the two Martian Trojans, 5261 Eureka
(e.g., Holt \& Levy 1990; Mikkola et al. 1994) and 1998 VF31 (e.g.,
{\it Minor Planet Circular 33085}; Tabachnik \& Evans 1999), these
searches have been without success. In particular, Weissman
\& Wetherill (1974) and Dunbar \& Helin (1983) used photographic
plates, while Whiteley \& Tholen (1998) used CCD cameras, to look
unavailingly for terrestrial Trojans.  We are unaware of any
observational surveys of the Lagrange points of Venus, although some
tentative strategies for Mercurian Trojans have been suggested (e.g.,
Campins et al. 1996).

Recent technological advances have re-invigorated interest in this
problem. There are at least two promising lines of attack.  First,
very large format CCD arrays are now being built.  Wide-field CCD
imaging has already transformed our knowledge of the outer Solar
System with the discoveries of over two hundred faint trans-Neptunian
or Kuiper-Edgeworth belt objects (e.g., Luu \& Jewitt 1998; Jewitt \&
Luu 1999). The same strategies may well bear fruit in the inner Solar
System, with the Lagrange points of the terrestrial planets obvious
target locations. A second way of finding asteroids in the inner Solar
System is by exploiting the large databases to be provided by
satellites like {\it The Global Astrometry Interferometer for
Astrophysics} (GAIA, see
``http://astro.estec.esa.nl/SA-general/Projects/GAIA/gaia.html'').
Ground-based surveys like {\it The Sloan Digital Sky Survey} (SDSS,
see ``http://www.sdss.org/'') will also be useful for objects exterior
to the Earth's orbit.

\begin{table}
\begin{center}
\begin{tabular}{|c|c|c|c|c|} \hline
Planet & $N$ & $i$ & $\lambda$ & $a/\aplanet$ \\ \hline
Mercury & $583$ & $0^\circ-10^\circ$ & $325^\circ-35^\circ$ 
& $0.9988-1.0012$ \\ \hline
Venus & $2000$ & $0^\circ-16^\circ$ & $15^\circ-345^\circ$
& $0.996-1.004$ \\ \hline
Earth & $2000$ & $0^\circ-40^\circ$ & $10^\circ-350^\circ$ 
& $0.994-1.006$ \\ \hline
Mars & $500$ & $8^\circ-40^\circ$ & $30^\circ-130^\circ$ 
& $1.000$ \\ \hline
Mars & $500$ & $10^\circ-40^\circ$ & $230^\circ-330^\circ$ 
& $1.000$ \\ \hline
\end{tabular}
\end{center}
\caption{This shows the number of test particles for each planet.
The initial conditions are drawn uniformly from the ranges in
inclination $i$, differential longitude $\lambda$ and semimajor axis
$a$ shown.  These test particles are integrated for 1 million years to
provide the survivors recorded in Table 2.}
\label{table:tableoflimits}
\end{table}
Motivated by these observational opportunities, this paper presents
the observable properties of numerically long-lived clouds of
asteroids orbiting about the Lagrange points of the terrestrial
planets.  Our clouds are generated from our earlier numerical
integrations of test particles for timescales up to 100 million years
(Tabachnik \& Evans 2000, henceforth Paper I). These are the longest
currently available integrations, although of course they span a small
fraction of the estimated 5 Gyr age of the Solar System. The survivors
from our earlier integrations are re-simulated and re-sampled to
provide the distributions of observables. These include the contours
of probability density on the sky, as well as the distributions of
proper motions and the brightnesses.

The paper discusses our strategy for constructing the observable
properties of clouds of coorbiting asteroids in Section 2.  Venus, the
Earth and Mars are examined in Sections 3, 4 and 5. The case of
Mercury, the hardest of all, is postponed to Section 6. For each
planet, we discuss both the synthetic observations and the optimum 
search strategy.
\begin{table}
\begin{center}
\begin{tabular}{|c|c|c|c|c|} \hline
Planet & Survivors & Trojans at $\Lfour$ 
& Trojans at $\Lfive$ \\ 
& (Sample I) & (Sample II) & (Sample II) \\ \hline
Mercury &$321$ & - & - \\ \hline
Venus & $1691$ & $359$ & $346$ \\ \hline
Earth & $982$ & $227$ & $202$ \\ \hline
Mars & $909$ & $487$ & $422$ \\ \hline
\end{tabular}
\end{center}
\caption{This shows the number of survivors around the Lagrange points
of each planet remaining after the 1 Myr integration. Sample I comprises
all the stable orbits, whether horseshoes or tadpoles. Sample II is
just the tadpole orbits.}
\label{table:tableoforbits}
\end{table}

\begin{figure*}
\centerline{\psfig{figure=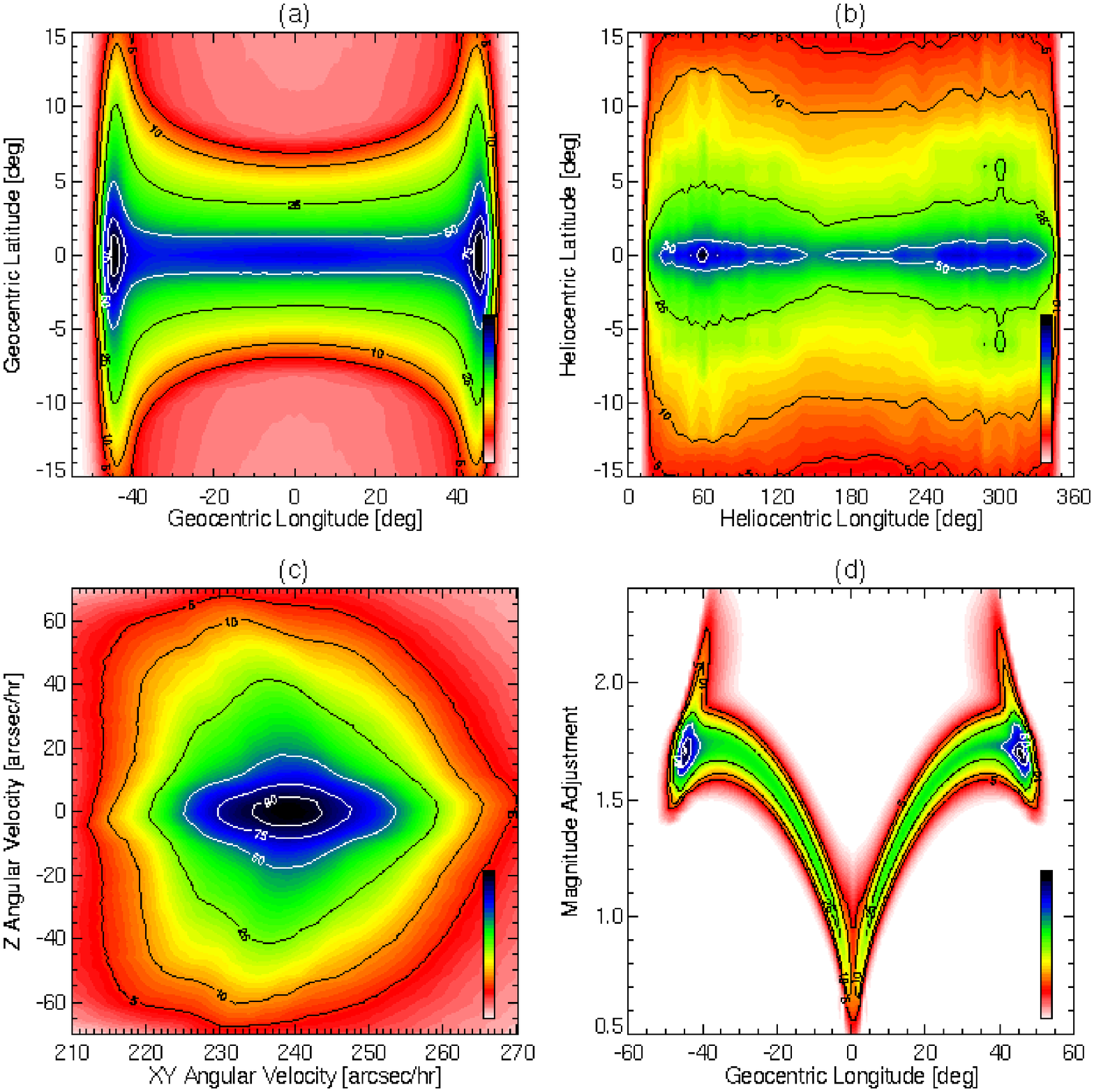,height=0.8\hsize}}
\caption{This shows the observable properties of Sample I of
long-lived horseshoe and tadpole orbits around the Venusian Lagrange
points. The panels illustrate the probability density in the planes of
(a) geocentric ecliptic latitude and longitude, (b) heliocentric
ecliptic latitude and longitude, (c) angular velocity in the ecliptic
($X-Y$) versus angular velocity perpendicular ($Z$) to the ecliptic,
and (d) magnitude adjustment versus geocentric longitude. In this and
subsequent figures, the colour scale ranges from deep blue ($90\%$
contour) through green ($50\%$) to red ($10\%$).}
\label{fig:venuspanelsall}
\end{figure*}
\begin{figure*}
\centerline{\psfig{figure=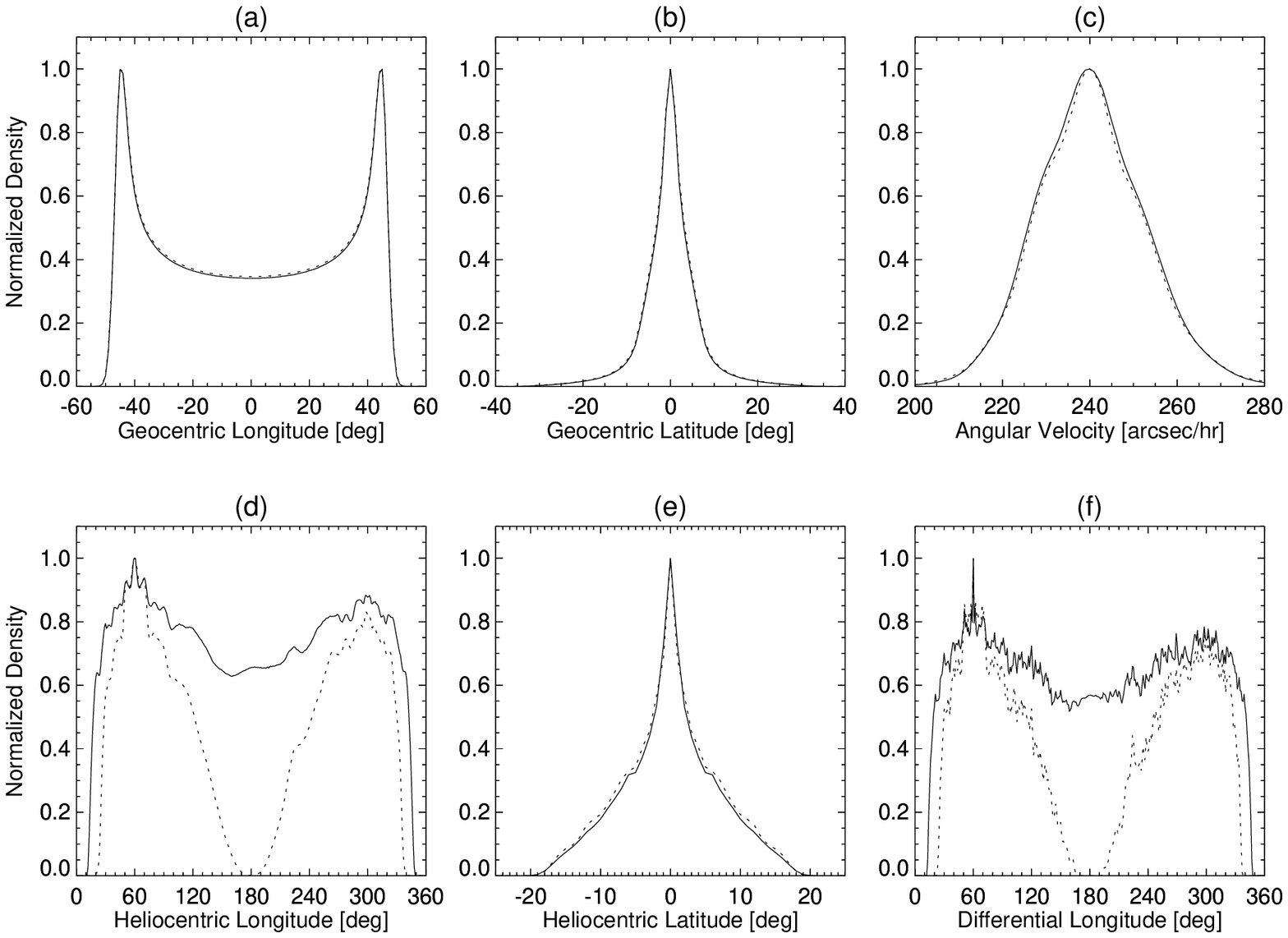,height=0.55\hsize}}
\caption{The one-dimensional probability distributions for synthetic
observations of coorbiting Venusian asteroids. These are (a)
geocentric longitude (b) geocentric latitude, (c) angular velocity,
(d) heliocentric longitude, (e) heliocentric latitude and (f)
differential longitude. Full lines represent Sample I, dotted lines
Sample II.}
\label{fig:venuslines}
\end{figure*}

\section{Simulations and Synthetic Observations}

With the results of Paper I in hand, it is straightforward to build
initial populations of stable test particles coorbiting with each of
the terrestrial planets.  For Venus, the Earth and Mars, Figures 6,
10, 11, 13 and 16 of Paper I are used to provide limiting values for
the semimajor axes, differential longitudes and inclinations, as
listed in Table~\ref{table:tableoflimits}.  The intervals are
replenished uniformly with random values in order to create sets of
2000 initial conditions in the case of Venus and the Earth, 1000 in
the case of Mars. In all cases, initial eccentricities, longitudes of
ascending node and mean anomalies coincide with the parent
planet. This collection of test particles is integrated for 1 Myr and
any test particles that enter the sphere of influence of a planet or
become hyperbolic are removed. Therefore, the number of objects in the
final samples vary from planet to planet and is recorded in
Table~\ref{table:tableoforbits}. For Venus and the Earth, results are
presented for two samples. The first (henceforth Sample I) comprises
stable tadpole and horseshoe orbits, the second (Sample II) only
tadpole or truly Trojan orbits. For Mars, the entire sample is tadpole
orbits, as no horseshoes survived the integrations of Paper I. (The
case of Mercury is different and is discussed in Section 6).

Once the population of long-lived test particles has been generated,
they are integrated again for 1 Myr. The positions and velocities are
then sampled every 2.5 yr in various reference frames and several
coordinate systems to generate our synthetic observations. In
particular, it is useful to present results in geocentric ecliptic
longitude and latitude ($\ellg, \betag$) in which the vantage point is
the Earth and the Sun is in the direction ($\ellg=
0^\circ,\betag=0^\circ$). The location of the $\Lfour$ and $\Lfive$
points in geocentric ecliptic coordinates differs from planet to
planet.  Sometimes, we also use heliocentric ecliptic latitude and
longitude coordinates ($\ellh, \betah$), in which the Sun is at the
origin and the planet under scrutiny lies in the direction
($\ellh=0^\circ, \betah=0^\circ$).  Now the leading $\Lfour$ cloud is
always roughly centered on ($\ellh = 60^\circ,\betah = 0^\circ$) and
the trailing $\Lfive$ cloud at ($\ellh =
300^\circ,\betah=0^\circ$). The angular velocities in the plane of,
and perpendicular to, the ecliptic are also sampled.  It is useful to
know the distribution of velocities as this determines the observing
strategy.  Finally, it is also helpful to know where the brightest
asteroids lie. This is of course controlled by the position, size and
albedo of the asteroid. The latter two quantities are unknown, but the
effects of the change of brightness with position can be investigated
by monitoring the magnitude adjustment $\magad$ (c.f., Wiegert,
Innanen \& Mikkola 2000). This is the difference between the apparent
magnitude $m_V$ and the absolute magnitude $H_V$ for each test
particle. In other words,
\begin{equation}
\magad = m_V (\alpha,r, \Delta)-H_V,
\end{equation}
where $\alpha$ is the solar phase angle (the angle subtended by the
Sun-asteroid-Earth system), while $r$ and $\Delta$ are the
heliocentric and geocentric distances of the asteroid.  Using the IAU
two-parameter magnitude system (Bowell et al. 1989), the magnitude
adjustment is
\begin{equation}
\magad = -2.5\log[(1-G)\Phi_1(\alpha) + \Phi_2(\alpha)] + 
5 \log r\Delta.
\end{equation}
Here, we have defined
\begin{eqnarray*}
\textrm{where} \; \;\left\{ \begin{array}{l} 
\Phi_i = W\Phi_{\rm{S}i} + (1-W)\Phi_{\rm{L}i};
\;\;\; i=1,2,\\ 
W=\exp \left[-90.56 \, \tan^2 \frac{1}{2} \alpha
\right],\\
\Phi_{\rm{S}i} = 1-\frac{\displaystyle C_i \sin
\alpha}{\displaystyle 0.119+1.341 \sin \alpha - 0.754 \sin^2 \alpha},\\
\Phi_{\rm{L}i} = \exp \left[-A_i \left( \tan \frac{1}{2}
\alpha \right)^{B_i} \right],\\
\end{array} \right.
\end{eqnarray*}
with the numerical constants such that $A_1 = 3.332, A_2 = 1.862, B_1
= 0.631, B_2 = 1.218, C_1 = 0.986$ and $C_2 = 0.238$. The slope
parameter $G$ is set to 0.15 corresponding to low-albedo C-type
asteroids. C-type are the most common variety of asteroids
constituting more than 75\% of the Main Belt. They are extremely dark
with an albedo of typically $0.03$. They are chemically similar to
carbonaceous chondritic meteorites (so they have approximately the
same composition as the Sun, minus hydrogen, helium and other
volatiles).

If the asteroid is seen at zero phase angle at a position where both
$r$ and $\Delta$ are 1 AU, then its magnitude adjustment vanishes. The
formula for the magnitude adjustment ceases to hold when the phase
angle exceeds $120^\circ$. This happens occasionally in our
simulations (for example, for asteroids of the inferior planets).
These cases are not included in our synthetic observations. They are
quite unimportant from an observational point of view, as the
asteroids then lie too close to the Sun to be detected in the visible
wavebands. The most useful datum appears to be the average magnitude
adjustment at the regions of greatest concentration of coorbiting
asteroids on the plane of the sky. This is calculated for the cloud
around each planet, by integrating the distribution of magnitude
adjustments at the heliocentric longitude corresponding to greatest
surface density.

We typically present our results using contour plots of the
probability density distributions in the plane of observables with the
colours from blue to red denoting regions from highest to lowest
density.  We draw contour levels representing points at $90\%, 75\%,
50\%, 25\%, 10\%$ and $5\%$ of the peak value. The contour maps always
show the unweighted synthetic observations and do not account for the
brightness of the test particles.  In addition to contour plots, we
also provide the one-dimensional distributions of geocentric and
heliocentric ecliptic longitude and latitude, differential longitude
and total angular velocity. The differential longitude is the
difference between the mean synodic longitudes of the test particle
and the planet. We have abstracted from our plots some statistics
which should prove useful for observational searches. These include
the means and the full-width half-maximum (FWHM) spreads of the
distributions, as well as 99\% confidence limits on the maximum and
minimum values. The statistics are given in
Tables~\ref{table:tableofstatv},~\ref{table:tableofstate},
~\ref{table:tableofstatm} and ~\ref{table:tableofstatme} for Venus,
the Earth, Mars and Mercury respectively.
\begin{table*}
\begin{center}
\begin{tabular}{|c|c|c|c|c|c|} \hline
Quantity & Sample & Mean & FWHM & Minimum & Maximum \\ \hline
XY Angular Velocity & I & 238.8\ah  & 28.2\ah  & 204\ah & 276\ah \\ \hline
Z Angular Velocity & I &0.0\ah  & 33.9\ah  & -68\ah  & 68\ah \\ \hline
Total Angular Velocity & I & 240.4\ah & 27.8\ah & 204\ah & 276\ah \\ \hline
Geocentric Latitude & I & $0.0^\circ$ & $6.2^\circ$ &
$-25^\circ$ & $25^\circ$\\ \hline
Geocentric Longitude & I & $45.0^\circ, -45.0^\circ$ 
& $10.4^\circ$ &$-48^\circ$ & $48^\circ$\\ \hline
Heliocentric Latitude & I & $0.0^\circ$ & $4.7^\circ$ &
$-17^\circ$ & $17^\circ$\\ \hline
Heliocentric Longitude & I & $60.0^\circ, 299.5^\circ$ 
& $325.3^\circ$ &$17^\circ$ & $343^\circ$\\ \hline
\end{tabular}
\end{center}
\caption{The statistical properties of the one-dimensional
distributions of observations for our sample of long-lived coorbiting
Venusian asteroids. The mean, full-width half-maximum (FWHM) and 99\%
confidence limits on the minimum and maximum values are recorded.}
\label{table:tableofstatv}
\end{table*}
\begin{figure}
\centerline{\psfig{figure=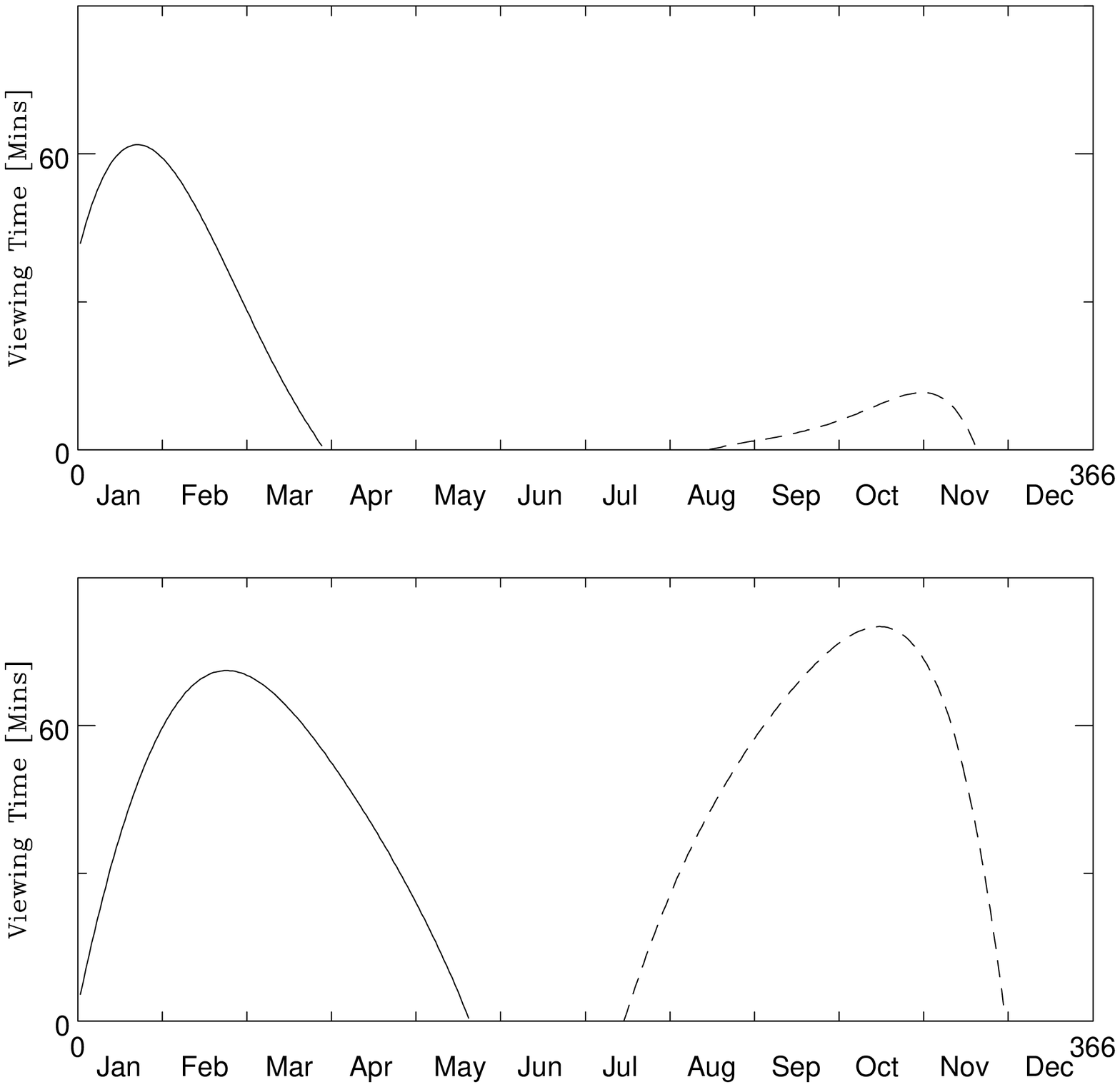,height=\hsize}}
\caption{This shows the observing time for the trailing (unbroken
line) and leading (broken line) Venusian Lagrange points from (a)
Mauna Kea and (b) La Silla observatories for the year 2000.}
\label{fig:venusview}
\end{figure}

\begin{table}
\begin{center}
\begin{tabular}{|c|c|c} \hline
Planet & $\magad$ & $R$ (in km) \\ \hline
Mercury & $0.6$ & $-$ \\ \hline
Venus & $1.7$ & $0.8$ \\ \hline
Earth & $2.0$ & $1.0$ \\ \hline
Mars & $3.5$ & $1.9$ \\ \hline
\end{tabular}
\end{center}
\caption{For the asteroidal cloud around each planet, this shows
the magnitude adjustment $\magad$ at the region of greatest
concentration on the sky. The GAIA all-sky survey satellite will scan
the triangular Lagrange points of Venus, the Earth and Mars.  The
magnitude adjustment is used to estimate the radius $R$ of the
smallest asteroid detectable by GAIA, assuming the survey is complete
to $V \approx 20$.}
\label{table:tableofgaia}
\end{table}

\section{Venus} 
\subsection{Observables}

\noindent
We first discuss the results for Sample I (all survivors).
Fig.~\ref{fig:venuspanelsall}(a) shows the enormous area of sky that
these objects can occupy. The 50\% contour encloses $\approx 300$
square degrees of sky. The distribution is centered around the
ecliptic. Note that the orbital plane of Venus, which marks the
midplane of the distribution of coorbiting objects, is misaligned with
the ecliptic by $\approx 3^\circ$. This accounts for some of the
thickness in the latitudinal direction. The highest probability
density occurs at geocentric ecliptic longitudes corresponding to
Venus' greatest eastern and western elongations. These are at $\ellg =
\pm \asin\, \aV \approx \pm 44^\circ$, where $\aV$ is the semimajor
axis of Venus in AU.  Fig.~\ref{fig:venuspanelsall}(b) shows the same
distribution in the plane of heliocentric ecliptic longitude and
latitude. The distribution is concentrated towards the plane of the
ecliptic, but the disk is quite thick.  There is an asymmetry evident
between the leading $\Lfour$ and trailing $\Lfive$ points.  For
$\Lfour$, the probability density has a narrow peak at the classical
value of ($\ellh =60^\circ, \betah=0^\circ$), but for $\Lfive$, the
peak is broader and runs from ($\ellh \approx 260^\circ - 310^\circ,
\betah = 0^\circ$). In fact, the Jovian Trojans are known to have a 
markedly asymmetric distribution with $\approx 80 \%$ librating about
the leading $\Lfour$ point. In all our simulations, asymmetries arise
because more of our initial sample of 2000 objects survive the
preliminary 1 million year integration about $\Lfour$ than $\Lfive$
(see Table~\ref{table:tableoforbits}). However, these asymmetries are
always rather mild.

The distribution of velocities in the plane of, and perpendicular to,
the ecliptic is shown in Fig.~\ref{fig:venuspanelsall}(c). The
velocity dispersion perpendicular to the ecliptic is much larger than
the velocity dispersion in the plane. Let us recall that the objects
in our sample come from a narrow range of initial semimajor axes
corresponding to the coorbital zone. The width of this zone can be
deduced from the resonance overlap criterion (Wisdom 1980). By itself,
this range in semimajor axes implies a very small scatter in the
velocities in the plane, so it is the distribution of eccentricities
that is responsible for most of the broadening. As regards the
velocity distribution out of the plane, the substantial scatter is
largely produced by the range of inclinations.  Finally,
Fig.~\ref{fig:venuspanelsall}(d) shows the distribution of magnitude
adjustment versus geocentric longitude. The brightest objects occur at
$\ellg \approx 0^\circ$. These are the asteroids at superior
conjunction. Even though they are furthest away from the Earth, this
is outweighed by the effects of the almost zero phase angle. Of
course, this portion is of little observational relevance. The
broadest range of magnitude adjustments occurs at the greatest eastern
and western elongations $\ellg \approx \pm 45^\circ$. Here, the phase
angle changes quickly for small changes in the longitude. At this
point, where the greatest concentration of objects occurs and so is
probably of most observational relevance, the typical magnitude
adjustment is $\approx 1.7$.

For the case of Venus, the contour plots for Sample II (Trojans alone)
are very similar to Sample I -- even down to the detailed shapes of
the contours. Accordingly, we do not present them here, although they
are available in Tabachnik (1999). The only figure that is markedly
different is the distribution as seen in the plane of heliocentric
longitude and latitude. Obviously, a sample of Trojans do not reach
the conjunction point at $\ellh = 180^\circ$ and so the probability
density falls to zero here.

Fig.~\ref{fig:venuslines} shows the one-dimensional distributions,
while Table~\ref{table:tableofstatv} gives the means, full-width
half-maximum (FWHM) thicknesses, the minima and maxima (at the 99\%
confidence level) of the distributions.  In each case, results are
presented for Sample I (full lines) and Sample II (dotted lines). Only
in the case of heliocentric ecliptic longitude and differential
longitude are there are any readily discernable differences between
Sample I and Sample II. The distributions in both heliocentric and
geocentric ecliptic latitude show that coorbiting Venusian asteroids
are most likely to be found in or near to the ecliptic. As regards
longitude, the distributions always peak at or close to the classical
Lagrange points, but they are broad. The horseshoe orbits in Sample I
cause the probability of discovery even at the conjunction points to
be far from insignificant.  The angular velocity of Venus is $240.4$
\arcsechr\ and this coincides exactly with the mean of the
distribution of total angular velocity. However, this distribution,
too, is rather broad with a FWHM of $27.8$ \arcsechr. This hints that
it may not be optimum to track the telescope with the expected rate of
Venus.
\subsection{Strategy}

\noindent
One strategy for finding coorbiting Venusian asteroids is wide-field
imaging.  Fig.~\ref{fig:venusview} shows the viewing time of the
Venusian Lagrange points as a function of calendar date for the year
2000, tailored for the case of the observatories at Mauna Kea (upper
panel) and La Silla (lower panel).  The figure is constructed by
taking the difference between the moment when the $\Lfive$ point
reaches 2.5 airmasses and the end of morning civil twilight. In the
later months of 2000, only the leading $\Lfour$ point is visible and
the figure gives the difference between the onset of evening civil
twilight and the moment when the $\Lfour$ point reaches 2.5 airmasses.
This is a very generous definition of observing time, but even so the
Venusian Lagrange points are visible in 2000 for, at best, barely more
than an hour.  In 2000, Venus is visible in the morning sky till the
middle of April; it reappears in the evening sky after early
August. In the early months of 2000, only the trailing $\Lfive$ point
is visible, whereas in the later months, only the leading $\Lfour$
point is.  Venus is below the plane of the ecliptic for all of 2000,
and so the southern observatories provide more propitious observing
conditions. For La Silla, the best times are February or March for the
$\Lfive$ point, and October or November for the $\Lfour$ point.
\begin{figure*}
\centerline{\psfig{figure=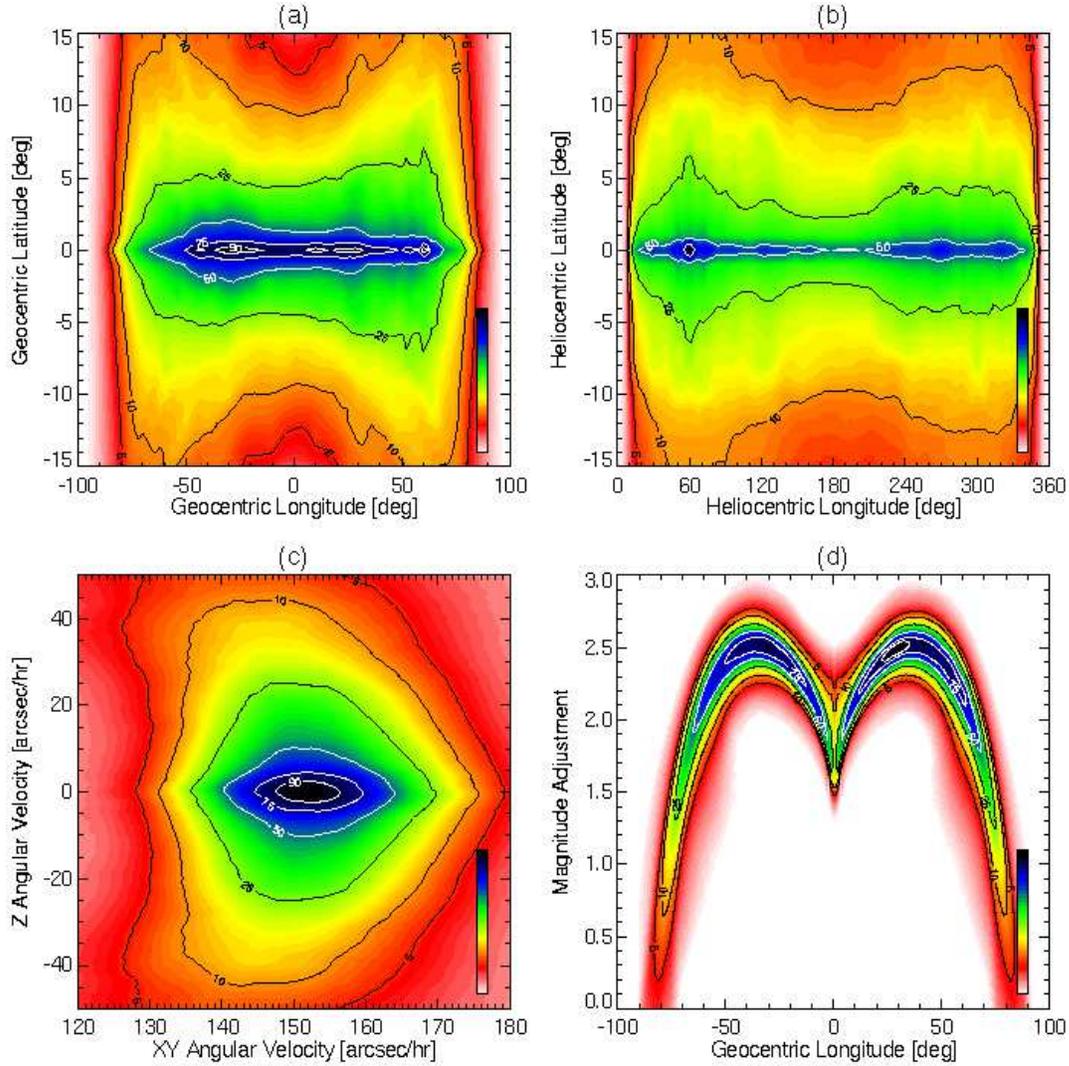,height=0.8\hsize}}
\caption{This shows the observable properties of Sample I of
long-lived horseshoe and tadpole orbits around the terrestrial
Lagrange points. The panels illustrate the probability density in the
planes of (a) geocentric ecliptic latitude and longitude, (b)
heliocentric ecliptic latitude and longitude, (c) angular velocity in
the ecliptic ($X-Y$) versus angular velocity perpendicular ($Z$) to
the ecliptic, and (d) magnitude adjustment versus geocentric
longitude.}
\label{fig:earthpanelsall}
\end{figure*}
\begin{figure*}
\centerline{\psfig{figure=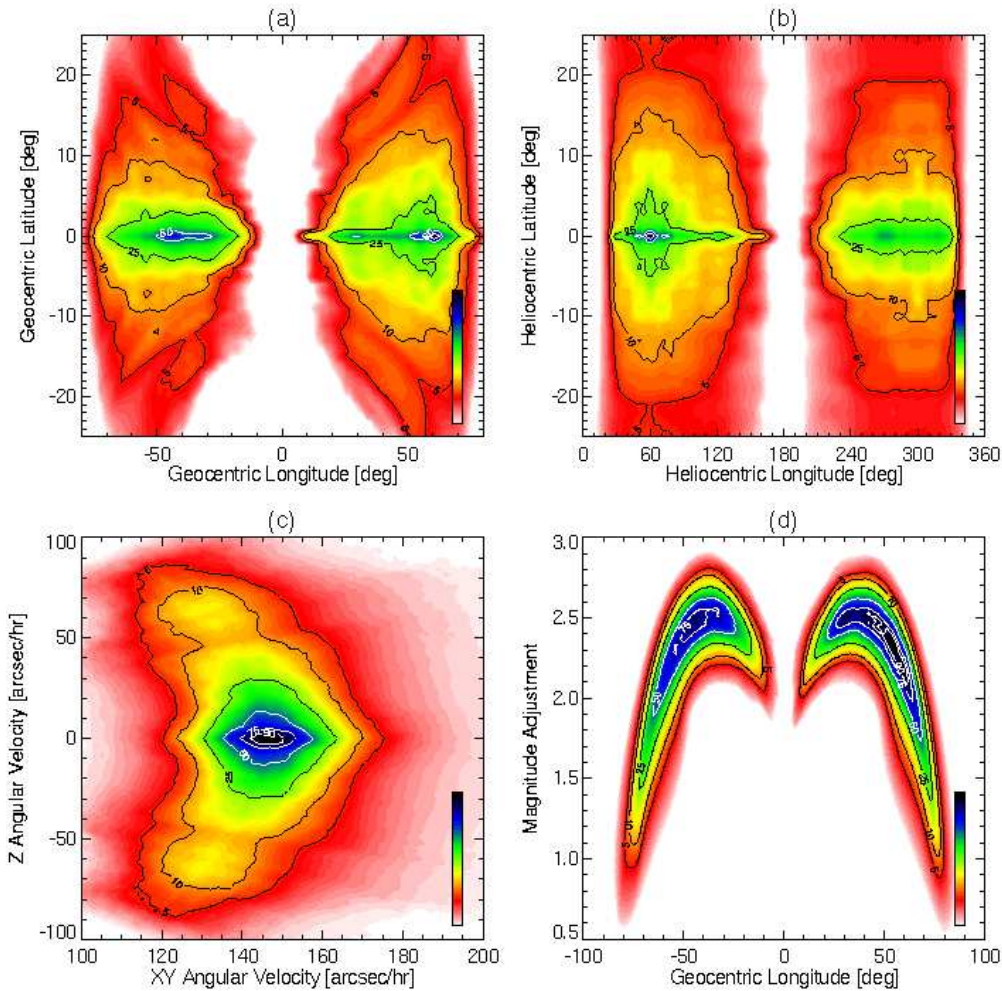,height=0.8\hsize}}
\caption{As Fig.~\ref{fig:earthpanelsall}, but for the sample II
(terrestrial Trojans alone).}
\label{fig:earthpanelstadpoles}
\end{figure*}
\begin{figure*}
\centerline{\psfig{figure=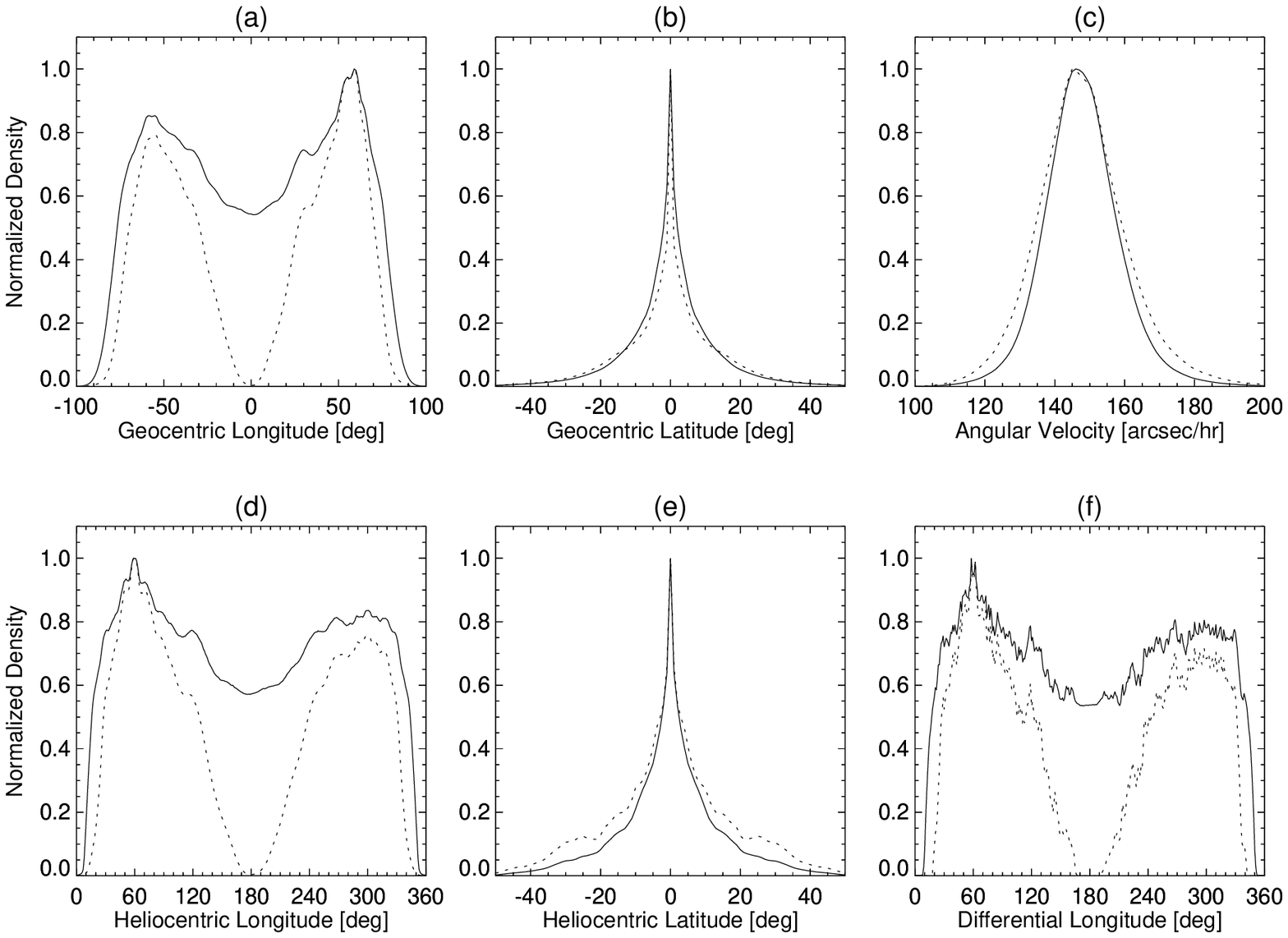,height=0.55\hsize}}
\caption{The one-dimensional probability distributions for synthetic
observations of coorbiting terrestrial asteroids. These are (a)
geocentric longitude (b) geocentric latitude, (c) angular velocity,
(d) heliocentric longitude, (e) heliocentric latitude and (f)
differential longitude. Full lines represent Sample I, dotted lines
Sample II.}
\label{fig:earthlines}
\end{figure*}

The other major problem is the very fast motions that these objects
typically have ($240.4$ \arcsechr\ is the mean for Venusian Trojans).
As the velocity distributions are broad, this may hinder techniques
such as tracking the camera at the average motion rate and looking for
unstreaked objects.  It may be better to track the telescope at the
sidereal rate and use short exposures to minimise trailing loss in
candidate Trojans. In any case, it is imperative that discovery of any
object happen preferably in real time, but certainly within 24 hours,
so that follow-up frames can be obtained as quickly as possible and
the candidate is not lost. This necessitates the development of
optimised software for real-time analysis (similar to that already
used in the Kuiper-Edgeworth Belt searches).

A second strategy for finding coorbiting Venusian asteroids is to
exploit the database that will be provided by GAIA. This is an ESA
all-sky survey satellite and will provide multi-colour and multi-epoch
photometry, spectroscopy and astrometry on over a billion objects. The
database will be complete to at least $V \approx 20$. GAIA will
observe to within $35^\circ$ of the Sun, so identification of
coorbiting asteroidal companions of Venus, the Earth and Mars will be
possible. {\it To our knowledge, this will be the first time the
Venusian Lagrange points have ever been scanned}. For each of the
planets, Table~\ref{table:tableofgaia} shows the minimum size C-type
asteroid that can be found with GAIA. For the Venusian cloud,
asteroids with radii $\gta 0.8$ km can be detected.  GAIA will observe
each asteroid roughly a hundred times over the mission
lifetime. Identification of successive apparitions of the same objects
in the database will be possible using estimates of the proper
motions.  Thanks to the astrometric accuracy of the GAIA satellite,
precise orbits for any newly discovered Trojans will be obtained.
\begin{table*}
\begin{center}
\begin{tabular}{|c|c|c|c|c|c|} \hline
Quantity & Sample & Mean & FWHM & Minimum & Maximum \\ \hline
XY Angular Velocity & I & 144.4\ah  & 22.5\ah  & 109\ah & 179\ah \\ \hline
Z Angular Velocity & I &0.0\ah  & 22.3\ah  & -84\ah  & 84\ah \\ \hline
Total Angular Velocity & I & 147.9\ah & 21.0\ah & 113\ah & 181\ah \\ \hline
Geocentric Latitude & I & $0.0^\circ$ & $4.0^\circ$ &
$-42^\circ$ & $42^\circ$\\ \hline
Geocentric Longitude & I & $59.6^\circ, -59.7^\circ$ 
& $153.4^\circ$ &$-84^\circ$ & $84^\circ$\\ \hline
Heliocentric Latitude & I & $0.0^\circ$ & $5.0^\circ$ &
$-42^\circ$ & $42^\circ$\\ \hline
Heliocentric Longitude & I & $60.0^\circ, 299.5^\circ$ 
& $328.1^\circ$ &$13^\circ$ & $347^\circ$\\ \hline \hline
XY Angular Velocity & II & 142.8\ah  & 28.3\ah  & 102\ah & 182\ah \\ \hline
Z Angular Velocity & II & 0.0\ah  & 27.3\ah  & -88\ah  & 88\ah \\ \hline
Total Angular Velocity & II & 147.9\ah & 25.1\ah & 108\ah & 186\ah \\ \hline
Geocentric Latitude & II & $0.0^\circ$ & $1.8^\circ$ &
$-43^\circ$ & $43^\circ$\\ \hline
Geocentric Longitude & II & $59.3^\circ, -59.3^\circ$ 
& $153.4^\circ$ &$-84^\circ$ & $84^\circ$\\ \hline
Heliocentric Latitude & II & $0.0^\circ$ & $5.19^\circ$ &
$-17^\circ$ & $17^\circ$\\ \hline
Heliocentric Longitude & II & $60.0^\circ, 299.5^\circ$ 
& $90^\circ$ &$27^\circ$ & $333^\circ$\\ \hline
\end{tabular}
\end{center}
\caption{The statistical properties of the synthetic observations of
our samples of long-lived coorbiting terrestrial asteroids. As usual,
Sample I is all surviving asteroids, Sample II is tadpoles only.}
\label{table:tableofstate}
\end{table*}
\begin{figure}
\centerline{\psfig{figure=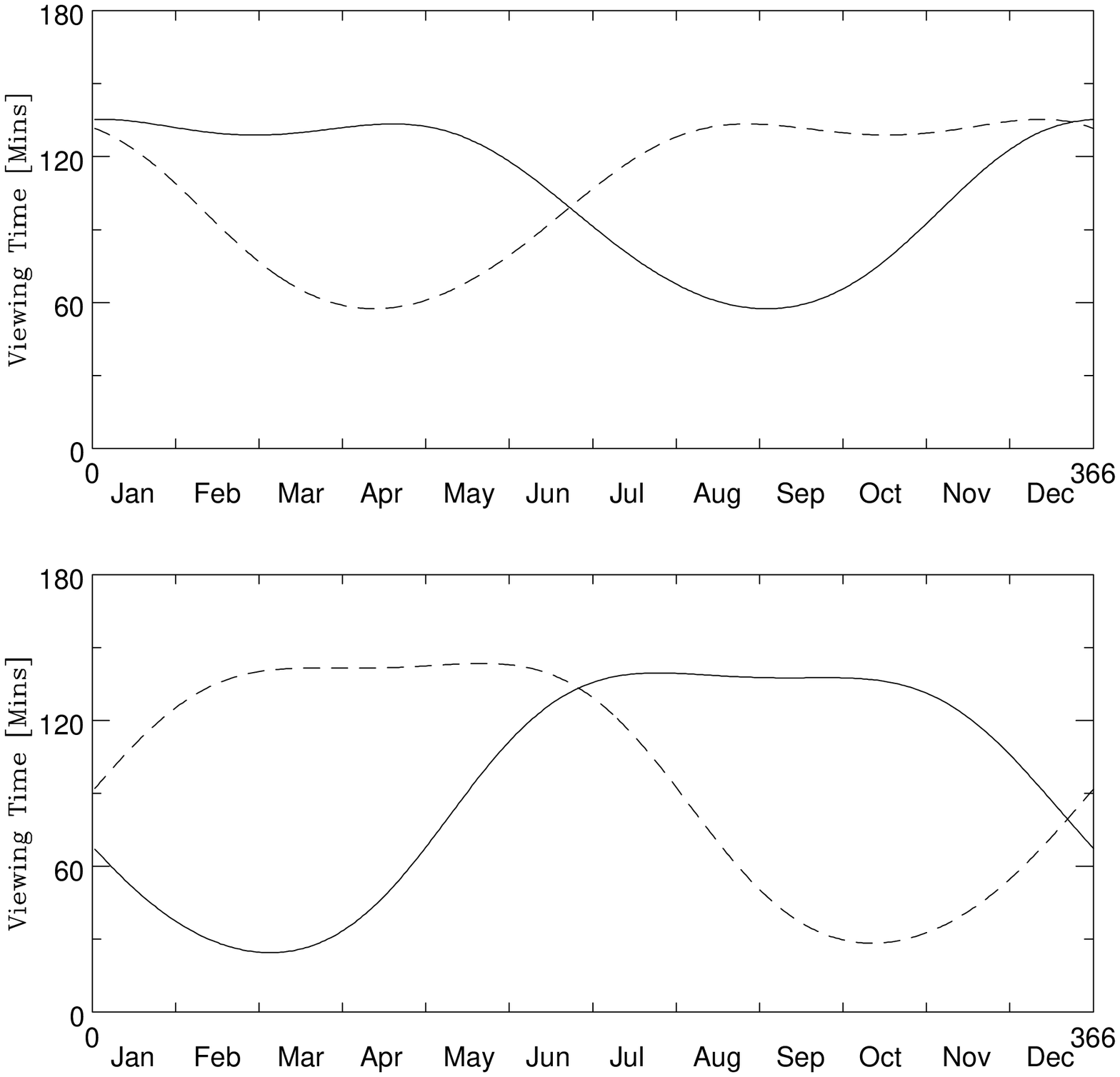,height=\hsize}}
\caption{This shows the observing time for the trailing (unbroken
line) and leading (broken line) terrestrial Lagrange points from (a)
Mauna Kea and (b) La Silla observatories for the year 2000.}
\label{fig:earthview}
\end{figure}
\section{The Earth}

\subsection{Observables}

\noindent
The Earth is known to have one coorbiting companion, 3753 Cruithne,
which follows a temporary horseshoe orbit (Wiegert, Innanen \& Mikkola
1997; Namouni, Christou \& Murray 1999). Figs.~\ref{fig:earthpanelsall} and
\ref{fig:earthpanelstadpoles} and shows the observable properties for
Samples I (tadpoles and horseshoes) and II (tadpoles only). There are
now more substantial differences between the two, and so we give the
contour plots for both the samples. The one-dimensional distributions
are given in Fig.~\ref{fig:earthlines} and some useful statistics in
Table~\ref{table:tableofstate}.  Our results can usefully be compared
to Wiegert et al. (2000), who also used numerical simulations to
derive observable properties of terrestrial Trojans.  Our approach
differs somewhat from theirs. First, they restricted their study to
tadpole orbits. Second, they used a number of plausible hypotheses to
set the semimajor axis, inclination and eccentricity limits on their
initial population, whereas we have used the results from our lengthy
numerical simulations in Paper I. Both approaches seem reasonable, and
the differences between our results give an idea of the underlying
uncertainties in the problem.

Fig.~\ref{fig:earthpanelsall}(a) shows the probability distribution in
the plane of geocentric ecliptic longitude and latitude for Sample
I. The $50 \%$ contour encompasses some 350 square degrees of sky, so
a huge area has to be surveyed for coorbiting asteroids. Even if the
search is restricted to tadpoles or Trojans,
Fig.~\ref{fig:earthpanelstadpoles}(a) shows that the $50 \%$ contour
encloses some 30 square degrees of sky.
Figs.~\ref{fig:earthpanelsall} and \ref{fig:earthpanelstadpoles} also
show the concentration of objects in the plane of heliocentric
ecliptic longitude and latitude.  As in the case of Venus, there is an
asymmetry between leading and trailing points. This is because more of
the initial population around the $\Lfour$ point survive the 1 million
year integration.  Of course, exact symmetry between $\Lfour$ and
$\Lfive$ is not expected because of differences in the planetary
phases. For neither Sample I nor Sample II does the greatest
probability density occur at the classical values of ($\ellg = \pm
60^\circ, \betag =0^\circ$), but at smaller geocentric ecliptic
longitudes. This effect was already noticed by Wiegert et al. (2000),
who argued that the highest concentration of Trojans in the sky occurs
at ($\ellg \approx \pm 55^\circ, \betag = 0^\circ$). The physical
reason for this is that test particles on both tadpole and horseshoe
orbits spend more time in the elongated tail.  Even so, the effect is
slight; the one-dimensional probability distributions displayed in
Fig.~\ref{fig:earthlines} show that, when integrated over all
latitudes, the maxima return to the classical values of $\ellg = \pm
60^\circ$.  Note that the thickness in the heliocentric ecliptic
latitudinal direction seems larger for the Venusian cloud than for the
terrestrial (compare Fig.~\ref{fig:venuspanelsall}(b) with
Fig.~\ref{fig:earthpanelsall}(b)). This seems surprising, as the
stable coorbiting satellites persist to much higher inclinations in
the case of the Earth as compared to Venus. The reason is geometric,
as the Earth is further from the Sun than Venus. In actual distances,
the FWHM thickness of the terrestrial asteroids is $\approx 0.09$ AU,
while for the Venusian asteroids, it is $\approx 0.06$ AU. This is as
expected, since the terrestrial distribution has a tail extending to
higher latitudes.

Figs.~\ref{fig:earthpanelsall}(c) and \ref{fig:earthpanelstadpoles}(c)
show the distributions generated by Samples I and II respectively in
the plane of velocities.  The Earth has a mean motion of $147.84$
\arcsechr, very close to the mean total angular velocity of both
Samples I and II. However, the distribution of velocities is broad,
with a FWHM of exceeding $20$ \arcsechr\ (see
Table~\ref{table:tableofstate}). Again, this calls into question the
applicability of the search strategies which take two or three
exposures of each field while tracking the telescope at a rate of
$150$ \arcsechr\ along the ecliptic (c.f., Whiteley \& Tholen 1999).
This is only effective if the distribution of expected longitude rates
is very narrow. If the coorbiting objects possess broad distributions
of angular motions, then, with this technique, not only are background
stars trailed, but so are the candidate objects (unless they are
moving at exactly the right rate). This renders detection very
difficult.

The magnitude adjustment distributions are shown in
Figs.~\ref{fig:earthpanelsall}(d) and
\ref{fig:earthpanelstadpoles}(d).  For Venus, the nearby asteroids are
among the faintest. The converse is true for the Earth. The nearby
objects with semimajor axes slightly greater than the Earth's benefit
both from small phase angles and small geocentric distances and so are
brightest (as also noted by Wiegert et al. 2000). The asteroids at the
Lagrange points in Sample I are on average $\approx 2.0$ magnitudes
fainter than their absolute magnitudes. This number changes only
slightly to $2.1$ magnitudes for the tadpole-only Sample II.  For the
same distribution of absolute magnitudes, the Venusian asteroidal
cloud is actually brighter than the terrestrial cloud by $\approx 0.3$
magnitudes at their respective regions of greatest concentration.

\subsection{Strategy}

\noindent
Observational searches for coorbiting terrestrial asteroids present
less severe difficulties than the case of Venus, though they are still
not easy.  Fig.~\ref{fig:earthview} shows the windows of opportunity
for the terrestrial Trojans at Mauna Kea (upper panel; c.f., Whiteley
\& Tholen 1998) and La Silla (lower panel). For the Earth's leading
Lagrange cloud, the figure is constructed by calculating the time
difference between the beginning of evening civil twilight and the
moment when the setting $\Lfive$ point reaches 2.5 airmasses. For the
Earth's trailing cloud, it is calculated by taking the difference
between the the moment when the rising $\Lfour$ point reaches 2.5
airmasses and then end of morning civil twilight. Observations can
proceed more leisurely than for the Venusian Lagrange points as,
during times of peak visibility, there are about 2 hours of viewing
time at both Mauna Kea and La Silla.  For La Silla, the best times are
February to June for the $\Lfive$ point, and August to November for
the $\Lfour$ point. 

Unlike Venus, the Earth's Lagrange points have already been the
subject of previous scrutiny. The most thorough survey to date is that
of Whiteley \& Tholen (1999), who scanned some 0.35 square degrees
near the $\Lfour$ and $\Lfive$ points to a limiting magnitude $R
\approx 22.8$ without success using a $2048 \times 2048$ mosaic
camera.  By comparison, let us recall from
Fig.~\ref{fig:earthpanelstadpoles}(a) that the 50\% contour encloses
some 30 square degrees for Sample II of Trojans alone.  Whiteley \&
Tholen tracked the camera at the average motion rate of $148$
\arcsechr\ for coorbiting terrestrial asteroids, taking generally two,
occasionally three, 300s exposures of each field. Assuming a typical
seeing of 0.8 arcsec, then any object with motions in the range
$\approx 143 - 153$ \arcsechr\ will move less than the seeing during
the two exposures. Only these objects will not appear streaked. Using
our velocity distributions in Fig.~\ref{fig:earthlines}, we estimate
that such objects comprise only about 35\% of our sample.  The actual
efficiency of Whiteley \& Tholen's technique is harder to estimate, as
it depends on the details of its implementation, but it seems likely
that it is $\lta 50$\%.

There are also earlier photographic plate surveys, which covered
larger areas of sky, but with much less sensitivity. For example,
Dunbar \& Helin (1983) took sixteen Schmidt plates near the $\Lfour$
point, though only two near the $\Lfive$ point, with the 1.2m Palomar
telescope. They claim that the number of terrestrial Trojans with an
absolute magnitude $H_V \lta 20$ around $\Lfour$ must be $\lta
10$. Neglecting phase angle effects, the radius of an object given its
absolute magnitude $H_V$ is roughly
\begin{equation}
R \approx  10^{0.2(m_{V,\odot} - H_V)} \quad {\rm (in\,\, AU)}.
\end{equation}
Taking the apparent visual magnitude of the Sun $m_{V,\odot}$
as $-26.77$ (Lang 1980), this becomes
\begin{equation}
R \approx 3.8 \times 10^3 \Bigl( {0.03\over \pv} \Bigr)^{1\over2}
         10^{-0.2 H_V} \quad {\rm (in\,\, km)},
\end{equation}
where $\pv$ is the $V$ band geometric albedo. Let us assume $\pv =
0.03$, as is reasonable for dark C-type asteroids (e.g., Zellner
1979). Then, Dunbar \& Helin's (1983) claim is that there are at most
ten objects with $R \gta 0.4$ km around $\Lfour$. Suppose the number
of objects with radii greater than $R$ can be approximated by a power
law
\begin{equation}
N(>R) \propto R^{1-q}.
\label{eq:numberlaw}
\end{equation}
Taking $q \approx 3$, as suggested by studies of C-type main belt
asteroids in Zellner (1979), then $ N(>0.1\,\km) \approx 160$ around
$\Lfour$. It is clear that the observational constraints on the
existence of a few bright or many faint coorbiting terrestrial
asteroids are rather weak.
\begin{figure*}
\centerline{\psfig{figure=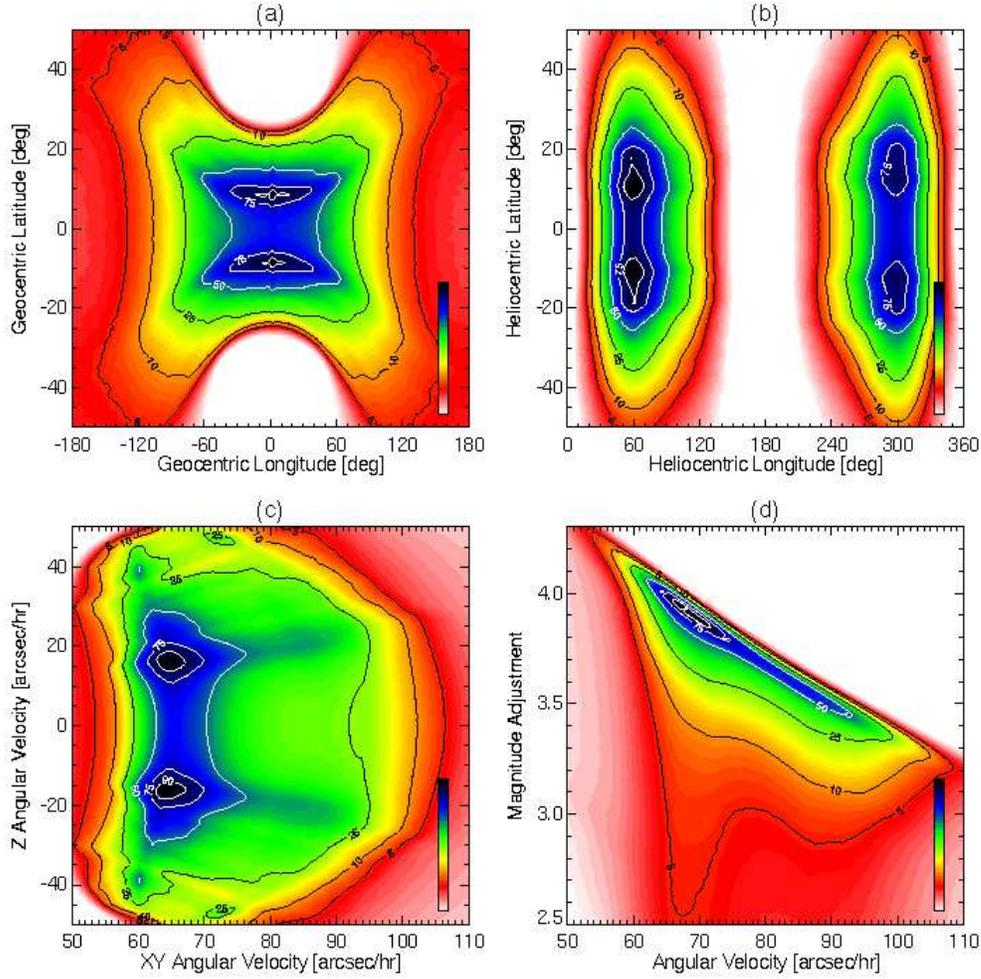,height=0.8\hsize}}
\caption{This shows the observable properties of Sample II of
long-lived tadpole orbits around the Martian Lagrange points. The
panels illustrate the probability density in the planes of (a)
geocentric ecliptic latitude and longitude, (b) heliocentric ecliptic
latitude and longitude, (c) angular velocity in the ecliptic ($X-Y$)
versus angular velocity perpendicular ($Z$) to the ecliptic, and (d)
magnitude adjustment versus geocentric longitude.}
\label{fig:marspanelsall}
\end{figure*}
\begin{figure*}
\centerline{\psfig{figure=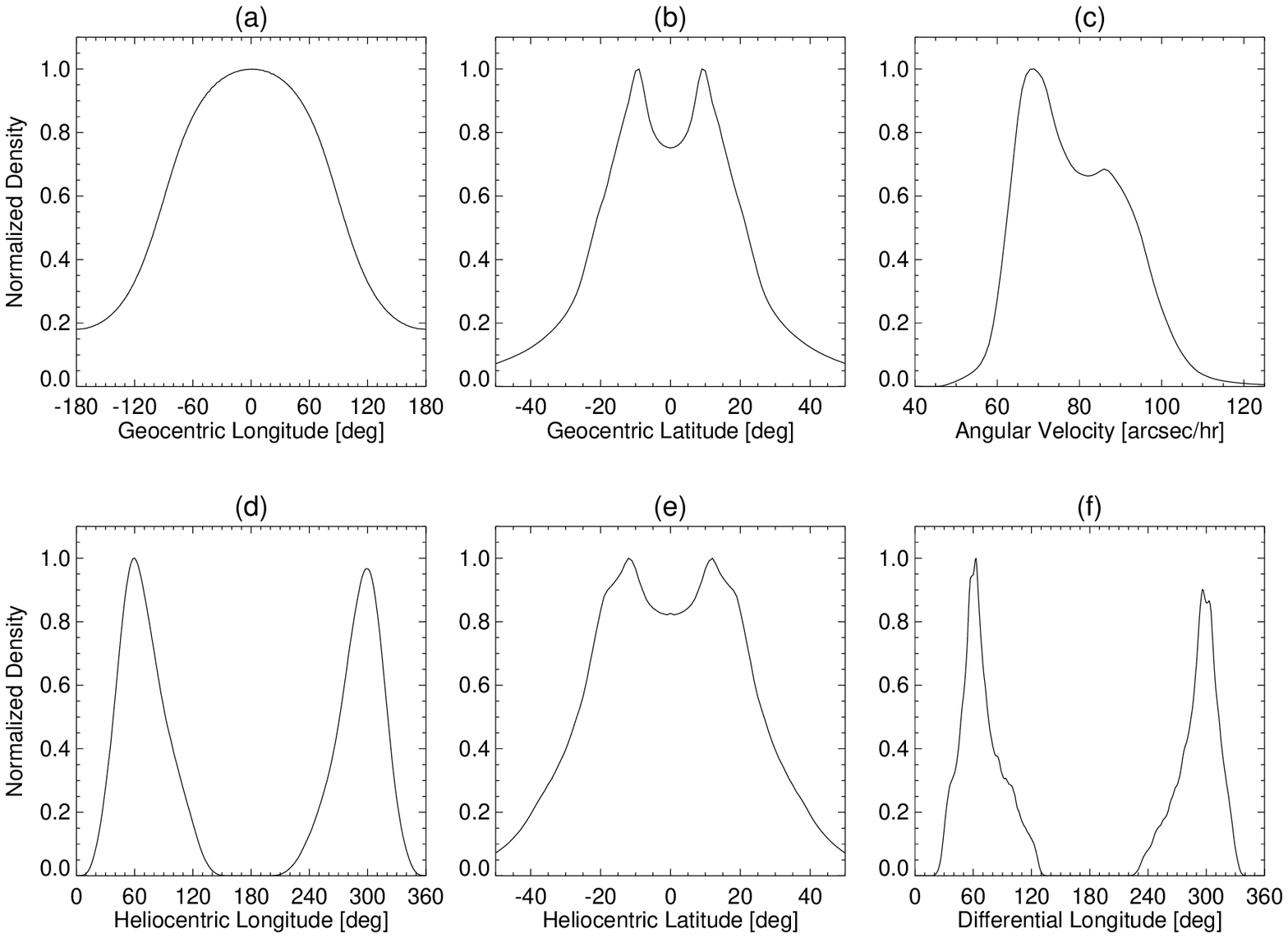,height=0.55\hsize}}
\caption{The one-dimensional probability distributions for synthetic
observations of coorbiting Martian asteroids. These are (a)
geocentric longitude (b) geocentric latitude, (c) angular velocity,
(d) heliocentric longitude, (e) heliocentric latitude and (f)
differential longitude.}
\label{fig:marslines}
\end{figure*}
\begin{table*}
\begin{center}
\begin{tabular}{|c|c|c|c|c|c|} \hline
Quantity & Sample & Mean & FWHM & Minimum & Maximum \\ \hline
XY Angular Velocity & II & 74.8\ah  & 27.0\ah & 52\ah & 116\ah \\ \hline
Z Angular Velocity & II & 0.0\ah  & 69.2\ah  & -51\ah & 51\ah \\ \hline
Total Angular Velocity & II & 78.8\ah & 32.2\ah & 54\ah & 114\ah \\ \hline
Geocentric Latitude & II & $0.0^\circ$ & $43.4^\circ$ &
$\lta -50^\circ$ & $\gta 50^\circ$\\ \hline
Geocentric Longitude & II & $-0.2^\circ$ 
& $197.7^\circ$ &$-180^\circ$ & $180^\circ$\\ \hline
Heliocentric Latitude & II & $0.0^\circ$ & $53.6^\circ$ &
$\lta -50^\circ$ & $\gta 50^\circ$\\ \hline
Heliocentric Longitude & II & $60.0^\circ, 299.5^\circ$ 
& $53.8^\circ, 49.7^\circ$ & $\lta 10^\circ$ & $\gta 350^\circ$\\ \hline
\end{tabular}
\end{center}
\caption{The statistical properties of the synthetic observations of
our samples of long-lived coorbiting Martian asteroids. Sample II is
tadpoles only.}
\label{table:tableofstatm}
\end{table*}
\begin{figure*}
\centerline{\psfig{figure=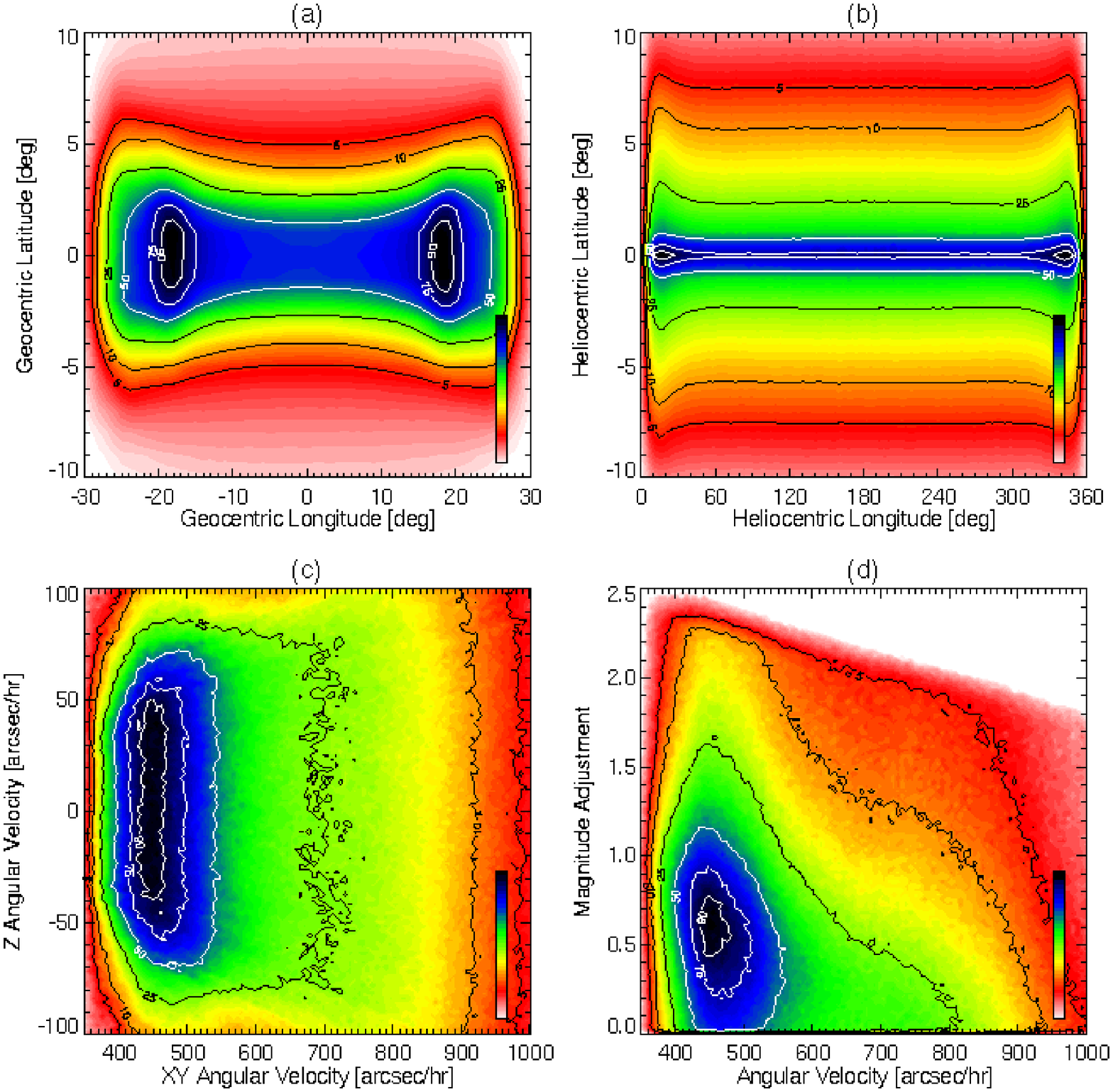,height=0.8\hsize}}
\caption{This shows the observable properties of Sample I of
long-lived horseshoe orbits around the Mercurian Lagrange points. The
panels illustrate the probability density in the planes of (a)
geocentric ecliptic latitude and longitude, (b) heliocentric ecliptic
latitude and longitude, (c) angular velocity in the ecliptic ($X-Y$)
versus angular velocity perpendicular ($Z$) to the ecliptic, and (d)
magnitude adjustment versus geocentric longitude.}
\label{fig:mercurypanelsall}
\end{figure*}
\section{Mars}

\subsection{Observables}

\noindent
Our sample of coorbiting Martian asteroids is composed solely of
inclined tadpole orbits. In Paper I, we found that asteroids in the
orbital plane of Mars are unstable and do not survive on 60 Myr
timescales. Without the range of semimajor axes provided by our
in-plane surveys, our procedure for generating initial conditions
gives only tadpoles.

Fig.~\ref{fig:marspanelsall}(a) and (b) shows that the greatest
concentration of Trojans lies above the ecliptic, at ($\ellg =
0^\circ, \betag \approx \pm 9^\circ$) or ($\ellh = 60^\circ, \betah
\approx \pm 12^\circ$). Interestingly, the two known Martian Trojans
(5261 Eureka and 1998 VF31) were discovered by opposition searches
close to the ecliptic. The discoveries were serendipitous, insofar as
the surveys were directed towards finding Near-Earth Objects. In fact,
the probability density in the ecliptic is typically $50 \%$ of the
peak. It is surely encouraging that search strategies that are far
from optimum have already yielded success.  Compared to our earlier
Figs~\ref{fig:venuspanelsall} and~\ref{fig:earthpanelsall} for Venus
and the Earth, the regions of high probability density extend to much
higher latitudes in the case of Mars. Of course, this is a consequence
of the fact that the stable Trojans survive only for starting
inclinations between $14^\circ$ and $40^\circ$ (see Paper I). As the
sample contains only tadpoles, the probability density falls to zero
at the conjunction points ($\ellh \approx 180^\circ$). In heliocentric
longitude, the greatest concentration lies close to the classical
Lagrange point values of $\ellh \approx 60^\circ$ and $300^\circ$. The
maximum in geocentric longitude occurs at $\ellg =0^\circ$. This is of
course a purely geometric effect, as the Martian asteroids are here
most distant from the Earth. Angular distances correspond to larger
true distances and so the sampling, which proceeds at equal time
instants, gives a higher concentration.
 
The mean motion of Mars is $78.6$ \arcsechr\ . However, as shown in
Fig.~\ref{fig:marspanelsall}(c), the maximum of the probability
distributions occurs at a velocity of $\approx 65$ \arcsechr\ in the
plane of the ecliptic and a velocity of $\approx \pm 15$ \arcsechr\
perpendicular to the ecliptic. Evident in the contour plot are wakes
of high concentration regions behind the twin peaks of the maxima,
which persist to quite high velocities. This structure is a
consequence of the unusual range of inclinations that long-lived
coorbiting Martian asteroids must possess.  The distribution of total
angular velocity is shown in Fig.~\ref{fig:marslines} and is
strikingly asymmetric. We have checked that the average of this
distribution is indeed the same as the mean motion of Mars, as it
should be. The largest velocities occurs when the test particles are
sampled near perihelion, the smallest velocities when the test
particles are sampled near aphelion.  Although there is a pronounced
peak in the probability distribution corresponding to the velocities
at aphelion, there is no peak corresponding to perihelion, but rather
a long tail. This happens because the sampling is at equal time
intervals and so -- as any orbit spends most time at aphelion -- this
feature is accentuated.  There is a mild secondary peak in the
one-dimensional distribution at $\approx 87$ \arcsechr. We suspect
that this is caused by the wakes and so is a consequence of the range
of inclinations (c.f. the case of Mercury).

The probability density is shown in the plane of magnitude adjustment
versus angular velocity in Fig.~\ref{fig:marspanelsall}(d). The
brightest objects are the ones closest to us because of the
advantageous distance and phase angle effects. They are found at
typical heliocentric ecliptic longitudes of $-20^\circ \lta \ellh \lta
20^\circ$.  The brightest objects are moving faster than average
(compare, for example, the $50 \%$ contours in
Fig.~\ref{fig:marspanelsall}(c) and (d)). At the regions of greatest
concentration ($\ellh = 60^\circ, \betah \approx \pm 12^\circ$), the
average magnitude adjustment is $\approx 3.5$.

The one-dimensional probability distributions are shown in
Fig.~\ref{fig:marslines}. The most interesting features are the
broadness of the distributions in velocity and in latitute. The FWHM
of the double-peaked latitudinal distributions recorded in
Table~\ref{table:tableofstatm} refers to the width about the
peaks. The total FWHM of these double-peaked distributions is of
course far broader, something like $\approx 60^\circ$. The
distribution in differential longitude may be compared with our
earlier presentation, using a different sample, in Tabachnik \& Evans
(1999). There, we found that the $\Lfive$ peak was larger than the
$\Lfour$, whereas here the reverse is the case. This cautions us
against reading too much into the mild asymmetries present in our
results. Stability arguments certainly can produce such slight
differences between $\Lfour$ and $\Lfive$. However, if there are any
large asymmetries in the data, it seems that these must be produced by
other mechanisms (such as planetary migrations).  Both the known
Martian Trojans librate about $\Lfive$, and it will be interesting to
see if this trend is maintained with further discoveries.

\subsection{Strategy}

\noindent
As Mars is a superior planet, the observational strategy for Trojan
searches is very considerably eased. The main reason why only two
Martian Trojans are known is that the optimum locations to scan are at
geocentric ecliptic latitudes of $\pm 9^\circ$ or, equivalently,
heliocentric ecliptic latitudes of $\pm 12^\circ$. Most survey work is
concentrated within a few degrees of the ecliptic itself. In fact,
both 5261 Eureka and 1998 VF31 were discovered when their orbits took
them close to the ecliptic (see Holt \& Levy 1990; {\it Minor Planet
Circular 33763}).

1998 VF31 was discovered by {\it The Lincoln Near Earth Asteroid
Research Project} (LINEAR, see ``http://
www.ll.mit.edu/LINEAR/''). LINEAR uses a $2560 \times 1960$ pixel CCD
camera to search to a limiting magnitude of $\approx 19.5$ for
Near-Earth Objects (NEOs). It has discovered over one hundred NEOs in
1999 alone. The main search strategy is to look near solar opposition
and around the ecliptic. LINEAR probably provides the best existing
constraints on numbers of objects at the Martian Lagrange points. The
Trojan 1998 VF31 has absolute magnitude $H_V = 17.1$ (see
``http://cfa-www.harvard.edu/iau/lists/MarsTrojans.html'').  From its
magnitude, 1998 VF31 has a radius of $\approx 1.4$ km.  Let us assume
that LINEAR has surveyed the Martian $\Lfive$ point complete to 17th
magnitude. This is probably not the case, but our assumption then
yields an underestimate for the total population. Assuming Poisson
statistics, then at the 99.9 \% confidence level, there can be at most
seven objects with $z \gta 1.4$ km at the $\Lfive$ point. Making the
same assumptions as for the terrestrial Trojans
(eq.~\ref{eq:numberlaw}), this suggests that the number of objects
with radii greater than 0.1 km orbiting around the Martian $\Lfive$
point is $\approx 1000$. Of course, caution should be exercised here,
as this is a crude estimate. LINEAR can go as deep as $V \approx
19.5$, which suggests that more Martian Trojans can be expected from
them soon.

There are two forthcoming surveys that may also provide Martian
Trojans. First, {\it The Sloan Digital Sky Survey} (SDSS) takes
multi-colour photometry in the \rdash, \idash, \zdash, \udash and
\gdash\ bands. It is not an all-sky survey, being limited to latitudes
well away from the Galactic plane ($|b| > 30^\circ$) and to ecliptic
longitudes around solar opposition ($100^\circ \lta \ellh \lta
280^\circ$).  Of the terrestrial planets, SDDS can only look for
Martian Trojans.  These may be provisionally identified by colours,
being bright in \rdash\ and \idash\ and faint in the other three
bands, together with an estimate of their proper motion.  If new
objects are identified in real time (or nearly so) with SDSS, then
immediate ground-based follow-up will permit accurate orbit
determination to ensure that they are not lost. Second, the space
satellite GAIA will provide an all-sky census of solar system objects,
complete to $V \approx 20$ and free from directional biases. This is
much needed. The most complete survey of the solar system remains that
of Tombaugh (1961), which covered a large fraction of the northern sky
to approximately 18th magnitude. Kowal's (1979) survey goes three
magnitudes deeper, but is concentrated around the ecliptic. Kowal's
main aim was to find slow-moving objects, so his survey does not
follow up objects with motions greater than 15 \arcsechr\ (like
Martian Trojans, amongst others). In other words, GAIA will give the
most complete catalogue of solar system objects out of the plane of
the ecliptic. It will be particularly valuable for objects like
Martian Trojans, whose distribution is biased out of the ecliptic.
Using our estimate of the average magnitude adjustment at the most
favoured location ($\ellh = 60^\circ, \betah \approx \pm 12^\circ$),
we reckon that GAIA will be sensitive to C-type Martian Trojans with
a radius of $1.9$ km or greater (see Table~\ref{table:tableofgaia}).

\begin{figure*}
\centerline{\psfig{figure=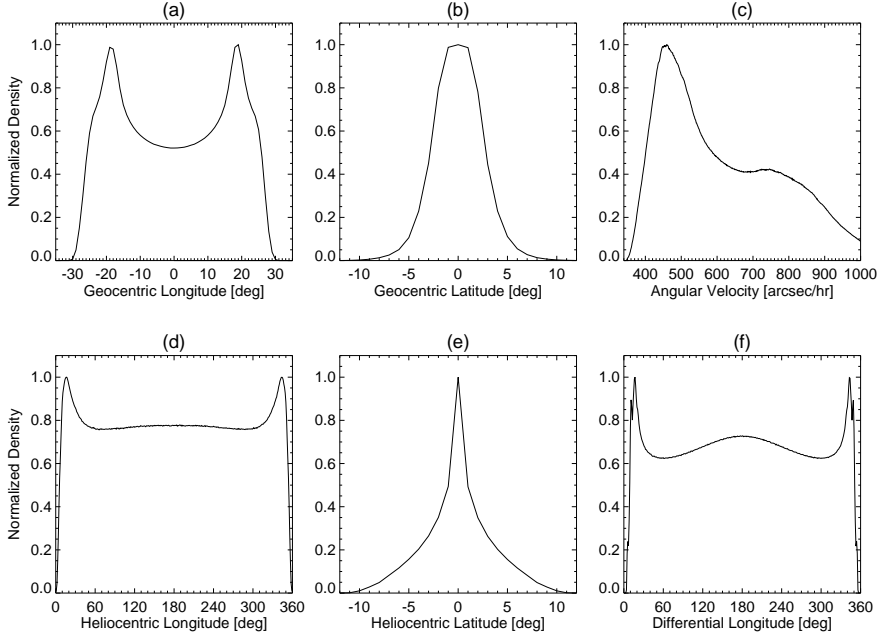,height=0.55\hsize}}
\caption{The one-dimensional probability distributions for synthetic
observations of coorbiting Mercurian asteroids. These are (a)
geocentric longitude (b) geocentric latitude, (c) angular velocity,
(d) heliocentric longitude, (e) heliocentric latitude and (f)
differential longitude.}
\label{fig:mercurylines}
\end{figure*}
\begin{table*}
\begin{center}
\begin{tabular}{|c|c|c|c|c|c|} \hline
Quantity & Sample & Mean & FWHM & Minimum & Maximum \\ \hline
XY Angular Velocity & I & 591 \ah  & 172.9\ah & 364\ah & 976\ah \\ \hline
Z Angular Velocity & I & 0.3\ah  & 170.4\ah  & $\lta -100$\ah & $\gta
100$ \ah \\ \hline
Total Angular Velocity & I & 459\ah & 186.5\ah & $\lta 350$ \ah & $\gta
1000$ \ah \\ \hline
Geocentric Latitude & I & $0.0^\circ$ & $5.7^\circ$ &
$-7^\circ$ & $7^\circ$\\ \hline
Geocentric Longitude & I & $18.5^\circ, -18.9^\circ$ 
& $51.5^\circ$ &$-28^\circ$ & $28^\circ$\\ \hline
Heliocentric Latitude & I & $0.0^\circ$ & $2.0^\circ$ &
$-9^\circ$ & $9^\circ$\\ \hline
Heliocentric Longitude & I & $16.0^\circ, 343.5^\circ$ 
& $348.5^\circ$ &$6^\circ$ & $354^\circ$\\ \hline
\end{tabular}
\end{center}
\caption{The statistical properties of the synthetic observations of
our samples of long-lived coorbiting Mercurian asteroids. Sample I is
horseshoes only.}

\label{table:tableofstatme}
\end{table*}
\section{Mercury}

\subsection{Observables}

\noindent
The case of coorbiting Mercurian asteroids is the most demanding of
all. The results of Paper I show that there is a long-lived population
of horseshoes with starting differential longitudes very close to
Mercury. This stable region is smaller than that of any of the other
terrestrial planets.

The procedure used to generate the synthetic observables in the case
of Mercury is different from the other three terrestrial planets.  The
sample of initial conditions is more strictly based on the results of
Paper I. The initial semimajor axis and differential longitudes are
determined from Figure 4 of Paper I.  Then, at each differential
longitude, an inclination is randomly chosen between $0^\circ$ and
$10^\circ$. The remaining three orbital elements are inherited from
the parent planet.  Of course, this set of initial conditions contains
a minor fraction of test particles which rapidly evolve on chaotic
orbits (eventually entering the sphere of influence of a planet or
becoming hyperbolic). So, we integrate the system for 1 Myr and then
retain only the bodies that survive. The reason for this difference in
procedure is that the stable zones of Figure 4 of Paper I are
particularly rich in structure, which we wished to retain as far as
possible.

The distributions of synthetic observables are shown in
Fig.~\ref{fig:mercurypanelsall}. There are a number of points of
interest. First, as is evident from
Figs.~\ref{fig:mercurypanelsall}(a) and (b), these objects are found
very close to the Sun. The greatest elongation of Mercury is at $\ellg
= \pm \asin \am = \pm 22.8^\circ$, where $\am$ is the semimajor axis
of Mercury in AU.  But, the maximum concentration of coorbiting
asteroids is at somewhat smaller longitudes, $\ellg = \pm
18^\circ$. Similarly, in heliocentric longitude, the maxima occur at
$\ellh = 16.0^\circ$ and $348.5^\circ$ and so are rather significantly
displaced from the classical values. The unfortunate effect of this is
to make the objects still harder to detect, as they are even closer to
the Sun than naive reasoning suggests.

Perhaps the most remarkable plot is Fig.~\ref{fig:mercurypanelsall}(c)
which shows broad distributions extending to very high velocities
indeed. The 99 \% confidence limits show that the velocities in the
plane of the ecliptic extends from $364$ \arcsechr\ to $976$
\arcsechr, whereas the velocities perpendicular to the ecliptic
stretch from $-100$ \arcsechr\ to $100$ \arcsechr. The highest
velocities of course occur at perihelion. The reason why such an
enormous range of velocities is possible is the eccentricity
fluctuations of Mercury, and hence the coorbiting population, during
the course of the simulation. Of course, the largest velocities occur
at perihelion for the most eccentric orbits. The one-dimensional
velocity distribution is shown in Fig.~\ref{fig:mercurylines}. Again,
there is a sharp peak corresponding to the velocities at aphelion,
whereas the higher velocities are spread out into a long tail. There
is a hint of a secondary maximum, although it is not as pronounced as
in the case of Mars. The properties of the velocity distribution also
conspire against discovery, as these very fast moving objects are
extremely hard to locate.  Fig.~\ref{fig:mercurypanelsall}(d) shows
the probability distribution in the plane of magnitude adjustment and
velocity.  The brightest objects occur close to the Sun, at geocentric
longitudes $\ellg \approx 0^\circ$. These are the asteroids close to
the conjunction point, which benefit from the almost zero phase
angle. A more relevant observational quantity is the typical magnitude
adjustment at the region of greatest concentration, which is $0.6$.

\subsection{Strategy}

\noindent
The Mercurian Lagrange points are always at more than 2.5 airmasses
between the hours of twilight, so that searches are not feasible in
the visible wavebands.  Given the difficulties, it comes as a surprise
to find any attempt has been made at this awkward problem. In fact,
both Leake et al. (1987) and Campins et al. (1996) have searched for
intra-Mercurial objects, the ``Vulcanoids'' (see Weidenschilling 1978;
Evans \& Tabachnik 1999), during the day in the infrared $L$
band. Campins et al. (1996) mention plans to look for Mercurian
Trojans explicitly. The proximity of coorbiting Mercurian asteroids to
the Sun causes them to be hot. The search strategy exploits the fact
that they have a substantial component of infrared emission.  At
$\approx 3.5$ microns, the background brightness of the daytime sky is
near its minimum, but the thermal emission from the Earth's atmosphere
is still comparatively low. This rationale suggests that the $L$ band,
or possibly the $K$ band, offer the best hopes for the searches.
Leake et al. (1987) searched between $9^\circ \le |\ellg | \le
12^\circ$ and $| \betag | < 1^\circ$. Their main motivation was to
find Vulcanoids rather than Trojans, and -- as can been seen from
Figs.~\ref{fig:mercurypanelsall} and~\ref{fig:mercurylines} -- this
location is not the region of highest concentration of coorbiting
asteroids. Nonetheless, the probability density is still significant,
being $\approx 60 \%$ of the peak. It would be worthwhile to repeat
Leake et al.'s (1987) experiment, but using fields of view centered on
$(\ellg = \pm 18^\circ, \betag = 0^\circ )$.

Since Leake et al.'s (1987) investigation, the physical size of
infrared arrays has increased a lot, but on telescopes the emphasis
has been on improving the sampling of the images rather than on
increasing the field of view. So, $256 \times 256$ InSb arrays are now
available but with fields of only about $80 \times 80$ arcsec (e.g.,
NSFCAM on {\it The Infrared Telescope Facility} (IRTF); see
``http://irtf.ifa.hawaii.edu/''). A number of larger arrays are just
being commissioned, but they tend to be Hg/Cd arrays with less
sensitivity in $L$ band than the In/Sb arrays.  NSFCAM has a limiting
magnitude in $L$ of 13.9; it is unclear whether this deep enough for
Mercurian asteroid searches.

There are also some relevant, older searches for faint objects close
to the Sun and for sun-grazing comets using solar eclipse photography
(Courten 1976). These yielded possible, but unconfirmed, detections of
minor bodies. These old searches extend down to approximately 9th
magnitude and cover fields of view of $15^\circ \times 15^\circ$
centered on the Sun.  A strategy exploiting more up-to-date technology
is to use space-borne coronagraphs. Although only a handful of
sun-grazing comets have ever been discovered from the ground, the
Large Angle Spectrometric Coronagraph (LASCO) on SOHO (see,
``http://sohowww.nascom.nasa.gov'') has already discovered several
hundred since beginning operation in 1995. Space-borne instruments are
evidently advantageous over ground-based for searches in the infrared,
as the problem of thermal emission from the Earth's atmosphere is
side-stepped. LASCO can detect objects that pass within a projected
distance of 30 solar radii, or $\approx 0.15$ AU. This corresponds to
geocentric ecliptic latitudes satisfying $|\ellg | \lta 9^\circ$. As
Fig.~\ref{fig:mercurypanelsall}(a) illustrates, the concentration of
coorbiting Mercurian asteroids is not negligible within this region, 
being $\approx 50\%$ of maximum

\section{Conclusions}

This paper has presented synthetic observations for long-lived clouds
of coorbiting asteroids around each of the terrestrial planets.  Our
earlier numerical integrations from Tabachnik \& Evans (2000) confirm
that some of these objects can survive for timescales up to 100 Myrs,
and probably longer still.  For Venus and the Earth, we have
investigated two samples -- one containing both tadpole and horseshoe
orbits, the other just tadpole or true Trojan orbits.  For Mars, our
sample contains only tadpoles; for Mercury, only horseshoes.  The
observable distributions in geocentric and heliocentric latitude and
longitude have been calculated, as well as their typical velocities
and magnitudes.  All these should be valuable aids in the
observational searches for these objects.  We briefly summarise our
main results for each planet in turn:

\medskip
\noindent
(1) For Venus and the Earth, the greatest concentration of objects on
the sky occurs close to the classical Lagrange points at heliocentric
ecliptic longitudes of $60^\circ$ and $300^\circ$. The distributions
are broad, especially if horseshoes are present in the sample.  The
full-width half maximum (FWHM) in heliocentric longitude for Venus is
$325^\circ$ and for the Earth is $328^\circ$.  In other words, truly
huge areas of sky have to be searched for these objects.  The mean and
most common velocity of these coorbiting satellites coincides with the
mean motion of the parent planet, but again the spread is broad with a
FWHM for Venus of $27.8$ \arcsechr\ and for the Earth of $21.0$
\arcsechr. The terrestrial Lagrange points have been searched with 
Schmidt plates and CCD cameras, but these searches cover only small
fields of view and rule out large bodies only. As the velocity
distributions of the coorbiting population are broad, one possible
strategy is to use wide field CCD imaging, tracking the camera at the
sidereal rate and using short exposures to minimise trailing loss in
the candidates. The alternative is to exploit the existing method of
Whiteley \& Tholen (1998) and track the camera at the likeliest motion
rate of the objects.  In 2000, the time available is at best about an
hour a night for the Venusian Lagrange points and about two hours for
the Earth. In this time, the CCD frames need to be trimmed, biased,
flat-fielded and searched for fast-moving objects, so that any
discovery takes place in real-time.

The Venusian Lagrange points appear never to have been scanned before,
although {\it The Global Astrometry Interferometer for Astrophysics}
(GAIA) will perform this task when it flies in the next decade. Using
the magnitude adjustments at the regions of greatest concentration on
the sky, we estimate that GAIA will identify coorbiting Venusian
asteroids with radii $\gta 0.8$ km, as well as terrestrial asteroids
with radii $\gta 1$ km.

\medskip
\noindent
(2) For Mars, our sample is composed solely of tadpoles alone.  The
greatest concentration of Trojans on the sky occurs at geocentric
ecliptic latitudes of $\pm 9^\circ$ or, equivalently, heliocentric
ecliptic latitudes of $\pm 12^\circ$, and does not occur in the
ecliptic. The reason why the distribution is distended to quite high
latitudes is that only inclined Martian Trojans are stable.  The most
likely velocity of the Trojans is $65$\ah, significantly less than the
Martian mean motion, while the FWHM of the velocity distribution is
$32.3$\ah.  The two already known Martian Trojans (5261 Eureka and
1998 VF31) were discovered serendipitously by searches for Near Earth
Objects. In fact, programs such as {\it The Lincoln Near Earth
Asteroid Research Project} (LINEAR) provide the best existing
constraints on the numbers of Martian Trojans. Assuming that LINEAR
has surveyed the Martian $\Lfive$ point down to a completeness of $V
\approx 17$, then we reckon that the number of undiscovered objects
with radii greater than $0.1$ km could still number several hundred.
Within the next decade, GAIA will provide an all-sky inventory of
solar system objects down to $V \approx 20$. It will be particularly
valuable for objects out of the plane of the ecliptic. We estimate
that Martian Trojans with typical radii of $\gta 1.9$ km will be
detectable by GAIA.

\medskip
\noindent
(3) For Mercury, our sample is composed solely of horseshoe
orbits. True Trojans, or tadpole orbits, are not present as they do
not survive on 100 Myr timescales. The most likely locations to find
coorbiting satellites are at geocentric longitudes of $\ellg \approx
\pm 18^\circ$, or heliocentric longitudes of $\ellh \approx
16.0^\circ$ and $348.5^\circ$. In other words, the most likely
longitudes for discoveries are significantly displaced from the
classical Lagrange points.  These objects can have velocities
exceeding $1000$ \arcsechr, although the most common velocity is $459$
\arcsechr, which is less than the Mercurian mean motion.  Any
ground-based searches for coorbiting Mercurian asteroids must take
place during the day and concentrate on the $L$ or possibly the $K$
bands.  The fields of view of infrared detectors are still quite small
and so this is a daunting proposition. Nonetheless, Leake et
al. (1987) covered two impressively large fields of view of 6 square
degrees on either side of the Sun back in 1987. Leake et al.'s
motivation was to look for Vulcanoids, and so their fields ( $9^\circ
\le |\ellg | \le 12^\circ$) are not optimally located for our
purposes. It would be interesting to repeat this experiment with
fields sited around $\ellg = \pm 18^\circ$.  Another hopeful strategy
is to use space-borne coronagraphs to search the area in the Solar
System around the Sun for minor bodies.

\section*{Acknowledgments}
NWE is supported by the Royal Society, while ST acknowledges financial
help from the European Community.  We wish to thank John Chambers,
Mark Lacy, Seppo Mikkola, Keith MacPherson, Prasenjit Saha and Scott
Tremaine for helpful comments and suggestions. Tim de Zeeuw and
Michael Perryman provided helpful information regarding GAIA, while
Scott Tremaine and Robert Lupton helped us with SDSS.  Above all, we
are grateful to Jane Luu for much wise advise on strategies for
observations and life.

\end{document}